\documentclass[aps,prl,twocolumn,color,psfig,epsf,superscriptaddress]{revtex4-2}
\usepackage[english]{babel}
\usepackage[T1]{fontenc}
\usepackage{tikz}
\usepackage{amssymb,amsmath,amsfonts}
\usepackage{textcomp}
\usepackage{braket}
\usepackage{nicematrix} 
\usepackage{graphicx,color}
\usepackage[dvipsnames]{xcolor}
\usepackage[utf8x]{inputenc}
\usepackage{bm}
\usepackage{mathbbol}
\usepackage{float}
\usepackage{comment}
\usepackage[linkcolor = blue, citecolor = red, urlcolor = blue, colorlinks = true]{hyperref}
\usepackage[version=4]{mhchem}

\usepackage{tabularx}
 
\begin{document}

\title{Optimal Translocation of Living \& Active Filaments in Confinement}

\author{M.~Vatin}
\affiliation{Department of Physics and Astronomy, University of Padova, Via Marzolo 8, I-35131 Padova, Italy.}
\affiliation{ICTP South American Institute for Fundamental Research, Instituto de Física Teórica, Universidade Estadual Paulista, São Paulo, SP, Brazil.}
\author{R.~C.~Sinaasappel}
\affiliation{van der Waals-Zeeman Institute, Institute of Physics, University of Amsterdam, 1098 XH Amsterdam, The Netherlands.}
\author{R.~Kamping}
\affiliation{van der Waals-Zeeman Institute, Institute of Physics, University of Amsterdam, 1098 XH Amsterdam, The Netherlands.}
\author{C.~Valeriani}
\affiliation{Departamento de Estructura de la Materia, Física Térmica y Electrónica, Universidad Complutense de Madrid, 28040 Madrid, Spain.}
\affiliation{GISC - Grupo Interdisciplinar de Sistemas Complejos, Universidad Complutense de Madrid, 28040 Madrid, Spain.}
\author{E.~Locatelli}
\affiliation{Department of Physics and Astronomy, University of Padova, Via Marzolo 8, I-35131 Padova, Italy.}
\affiliation{INFN, Sezione di Padova, Via Marzolo 8, I-35131 Padova, Italy}
\author{A.~Deblais}
\affiliation{van der Waals-Zeeman Institute, Institute of Physics, University of Amsterdam, 1098 XH Amsterdam, The Netherlands.}


\begin{abstract}
Active filament translocation through confined spaces is central to processes ranging from DNA transport through nanopores to cytoskeletal dynamics in cell migration. Here, we use living filamentous \textit{Tubifex tubifex} worms as a model system to investigate how activity and filament conformation govern transport in confinement. By tuning activity via temperature and tracking worm dynamics in a two-chamber geometry connected by a narrow bridge, we quantify their translocation behavior and conformational states.
In contrast to passive polymers and filaments, we find that contour length has negligible influence on trapping dynamics, while activity and reorientation jointly control escape. Strikingly, translocation efficiency is maximized at an intermediate temperature ($20^\circ$C), where a balance between directed propulsion and rotational diffusion optimizes exploration. We show that trapping times are governed by the interplay between the timescale of conformational rearrangements and the conformational entropy, quantifying the diversity of accessible shapes.
Simulations of tangentially driven active filaments quantitatively reproduce the experimental observations and provide a minimal physical framework to rationalize the existence of an optimal activity. More broadly, our results identify general principles governing active filament transport in confinement, with implications for both biological systems and the design of synthetic active slender objects.
\end{abstract}

\maketitle

{W}hen it comes to physical systems, polymers are among the best understood. From reptation theory to polymer translocation through nanopores, fundamental insights have enabled the prediction of a wide range of phenomena of practical importance, including polymer flow and transport in various contexts \cite{doi1986,fakhri2010,Locatelli2023}, as well as DNA sequencing via electric-field-driven nanopore translocation \cite{Sung1996,Muthukumar2001,Matysiak2006}.  

The situation changes dramatically when the filament itself is \textit{active}. Here, the interplay between activity and conformation gives rise to a plethora of emergent phenomena, including phase separation and novel mechanical responses \cite{sugi2019c,Deblais2020a,Deblais2020b,ozkan2021,Heeremans2022,patra2022collective,Patil2023,faluweki2023active,dunajova2023chiral}, which have no counterpart in \textit{passive} filaments. Understanding the non-equilibrium statistical physics of such systems remains an open challenge, both theoretically and experimentally. In the latter case, despite recent attempts \cite{Chao2025}, progress has been limited due to the lack of suitable and controllable model systems to investigate active filaments in controlled settings. 

Recently, living biological worms such as \textit{T.~tubifex} and \textit{L.~variegatus} have emerged as promising laboratory models for active filament physics \cite{Kudrolli2019,Deblais2020a,Deblais2020b,Heeremans2022,Deblais2023,Patil2023,Sinaasappe2025}. In nature, most active agents are elongated and deformable, similar to polymers, and their shape plays a crucial role in dictating both their interactions with the environment and their exploration strategies \cite{Mokhtari2019,kurzthaler2021,Fazelzadeh2023}. Shape is particularly relevant in confined spaces, where active filaments must bend, reorient, or deform to navigate through narrow constrictions \cite{xi2024,Sinaasappe2025}. Despite their ubiquity in biological and synthetic systems, the fundamental mechanisms governing active filament transport through even simple geometries, such as pores, channels and constrictions, remain poorly understood.

To investigate the transport properties of self-propelled, flexible filaments under confinement, we use \textit{T.~tubifex} worms, living biological filaments exhibiting active polymer-like behavior, as a model system. The worms are confined in a quasi-2D ``two-state'' geometry consisting of two 3D-printed circular chambers connected by a narrow bridge~\cite{bruckner2019}. This minimal setup enables the collection of a statistically significant dataset, allowing us to characterize their conformation, dynamics, and translocation properties over time.

We find that worms stochastically hop between cavities while efficiently reptating through the connecting bridge with occasional U-turns, due to reorientation. Trapping times in the cavities are not affected by the contour length, unlike in passive polymers driven by thermal fluctuations~\cite{muthukumar2003polymer,wong2008polymer,mohan2010polymer}. These observations are well captured by a tangentially driven active polymer model with homogeneous active forces~\cite{Bianco2018}, after calibration in unconfined conditions. Both experiments and simulations reveal a repeated alternation between compact conformations in the chambers and elongated ones in the bridge, enhancing transport via shape modulation. Finally, tuning the worm activity through the environmental temperature reveals a maximum in translocation efficiency at intermediate activity levels, corresponding to a maximum in the conformational entropy of the worms, \emph{i.e.}, the diversity of explored conformations. This optimum emerges from an interplay between translational, rotational diffusion and flexibility, which together facilitate cavity escape and bridge crossing.

\begin{figure*}
    \centering
    \includegraphics[width=2\columnwidth]{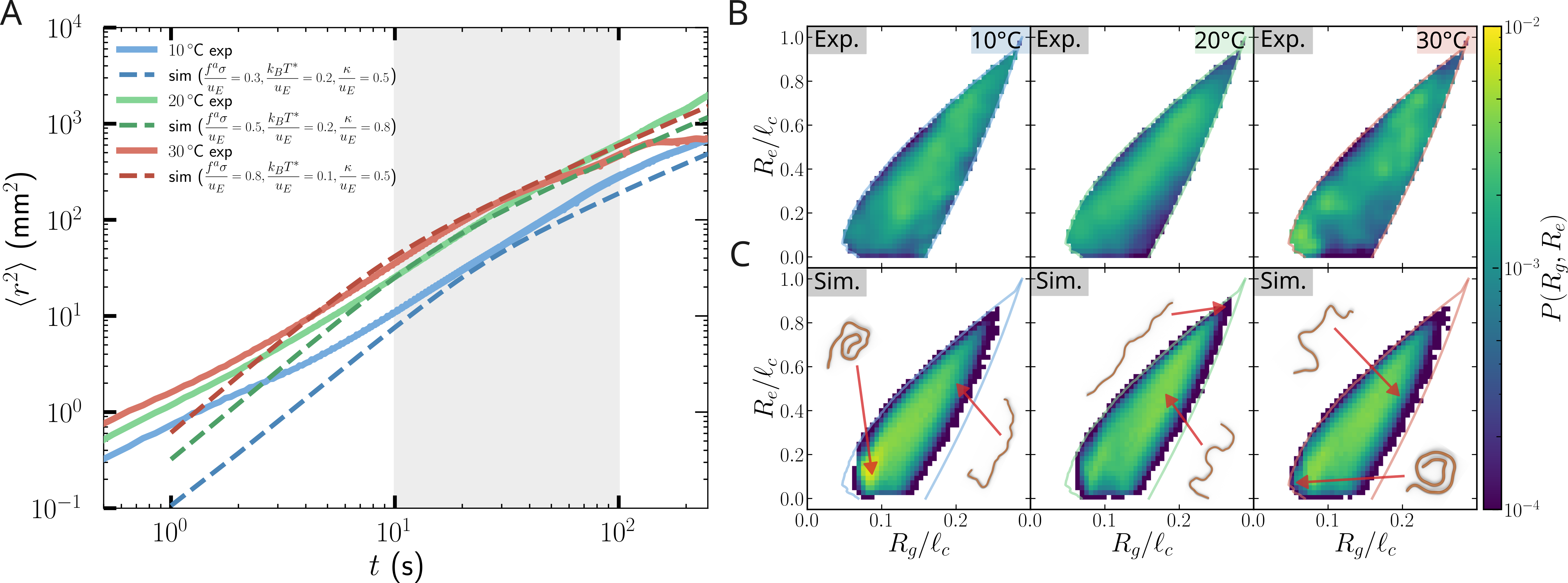}
\caption{\textbf{Living worm dynamics and conformations in free environments and parametrization of the active polymer model.}
(\textit{A})~Translational mean square displacement (MSD) of the center of mass of living worms in free environments (solid lines) at three different temperatures, together with the corresponding simulations (dashed lines). The input parameters of the tangentially driven polymer model (activity, effective temperature, and stiffness) are calibrated to reproduce the worm dynamics on the timescales relevant to the confinement experiments (gray shaded area). In the legend, $\sigma$ and $u_E$ denote the units of length and energy of the tangentially driven polymer model (see SI).
(\textit{B},\textit{C})~Conformational maps, where colors show the joint probability density in the normalized end-to-end distance $R_e/\ell_c$ versus normalized radius of gyration $R_g/\ell_c$ plane, with $\ell_c$ the contour length. (\textit{B}) Experimental measurements for living worms at temperatures $T = 10, 20, 30^\circ$C. (\textit{C}) Corresponding conformations obtained from simulations of the tangentially driven polymer model using the same parameters as in (\textit{A}). In both panels, solid lines delimit the region of accessible conformations: the upper bound corresponds to circular arcs of contour length $\ell_c$, with the rigid-rod limit at $R_g/\ell_c = 1/(2\sqrt{3})\simeq 0.29$ and $R_e/\ell_c = 1$, while the lower bound corresponds to the minimum-$R_g$ envelope extracted empirically from the simulation ensemble. Insets: representative snapshots of the active polymer conformations.}
	\label{fig:FreeEnvironement}
\end{figure*}

\subsection*{Conformation of living \textit{T.~tubifex} in free environments and calibration of the active filament model}

\begin{figure*}
\centering
\includegraphics[width=2\columnwidth]{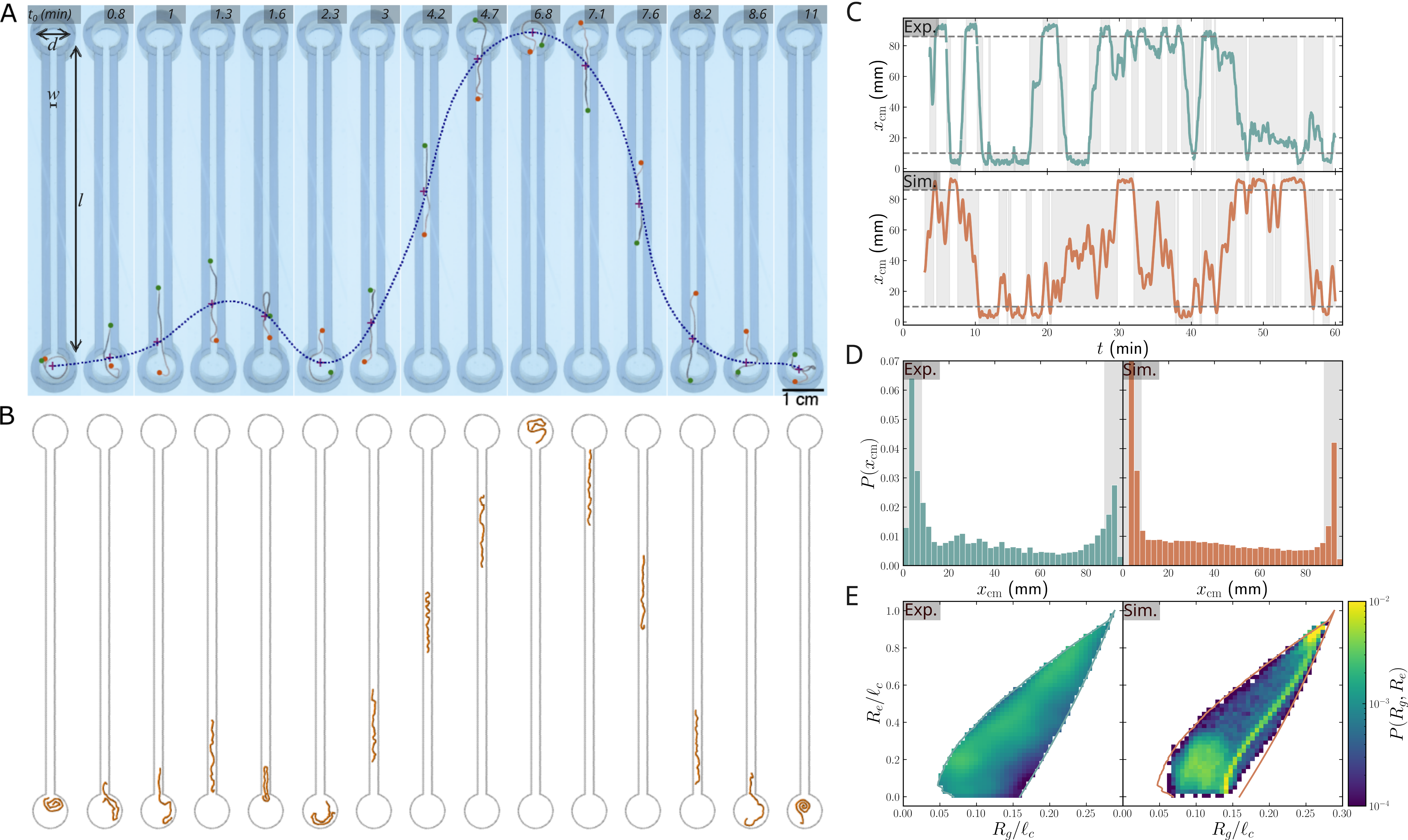}
\caption{\textbf{Translocation of active polymers.}
(\textit{A})~Time series of a living \textit{T.~tubifex} worm at ambient temperature ($T=20^\circ$C) in a quasi-two-dimensional channel composed of two circular chambers ($d = 8$~mm diameter) connected by a bridge ($w_b = 2$~mm, $L_b = 8$~cm). The worm is initially placed in one chamber; the head and tail (colored dots), overall contour conformation and center of mass (cross) are tracked over time.
(\textit{B})~Corresponding time series from simulations of an active polymer in the same geometry. Model parameters as in Fig.~\ref{fig:FreeEnvironement}\textit{A}. See Movies S1 and S2.
(\textit{C})~Center-of-mass position $x_{cm}$ as a function of time for a living worm (Top), and the simulated polymer (Bottom) during a typical $1$~h experiment. Gray shaded areas highlight the time spent in the cavities and in the connecting channel.
(\textit{D})~Probability density of $x_{cm}$: experimental data from ten worms of similar contour length (left) and corresponding simulations (right). Gray shaded areas indicate the chamber positions.
(\textit{E})~Conformational maps under confinement in the ($R_e/\ell_c$,$R_g/\ell_c$) plane, where colors represent the joint probability density estimated from frames sampled over the full 1-hour experiment (left: experiments, right: simulations).}
	\label{fig:Worm_Dynamics_in_Channel}
\end{figure*}

In a free environment, \textit{T.~tubifex} worms display stochastic wiggling motions that drive long-time diffusion of their center of mass (see Figure~\ref{fig:FreeEnvironement}\textit{A}), with a characteristic relaxation time $\tau_{e,0} \sim 10$~s, defined as the decorrelation time of the worm's end-to-end vector autocorrelation in free space~\cite{Deblais2020a,Deblais2020b,Sinaasappe2025}, at room temperature ($20^\circ$C). Their locomotion is effectively quasi-two-dimensional: being denser than water, worms sediment to the bottom substrate, although segments may occasionally lift out of plane.

To interpret these observations, we employ a minimal model of a tangentially driven active polymer~\cite{Winkler2020,Bianco2018,fazelzadeh2023effects}, which has been proven to capture the dynamics in free environment and simple settings~\cite{Sinaasappe2025,sinaasappel2026}. The filament is represented as a semiflexible bead-spring chain, with contour length comparable to that of the worms. Each bead evolves via overdamped Langevin dynamics and experiences a constant-magnitude active force $\mathbf{f}^a$, aligned with the local tangent of the polymer's contour. Stochasticity in the worm’s motion is captured through an effective thermal noise, while gravity is included as a downward force balanced by a planar wall, allowing for occasional out-of-plane excursions.

The model is thus described by three parameters: the active force magnitude $f^a$, the bending rigidity $\kappa$, and the effective temperature $T^*$. These parameters are calibrated by comparing simulations and experiments in the absence of confinement. Specifically, we minimize the difference between the experimental and simulated mean square displacement (MSD) of the center-of-mass on the relevant time scales of the experiments ([10--100] s, highlighted in grey in Figure~\ref{fig:FreeEnvironement}\textit{A}). The optimal parameter sets ($\kappa$, $T^*$, $f^a$) are thus determined for each environmental temperature; the corresponding simulations reproduce the experimental conformational statistics and the dynamics of isolated worms without further tuning (see Figure~\ref{fig:FreeEnvironement}\textit{B},\textit{C}). This calibration strategy avoids direct parameter extraction, which is often impractical in living systems, while ensuring that the model captures the intrinsic motility and flexibility of the worms. 
This benchmarking in the unconfined regime is essential: it establishes that the tangentially driven polymer model captures the behavior of \textit{T.~tubifex} worms before introducing geometrical constraints.

To characterize worm conformations, we extract the instantaneous centerline from videos and quantify the curvature profile, end-to-end distance $R_e$, and radius of gyration $R_g$. A custom tracking algorithm is used to reconstruct the worm contour and compute these observables, together with the center-of-mass position $x_{cm}$ (see Fig.~S1 and S2, and SI for details).

Figure~\ref{fig:FreeEnvironement}\textit{B} shows the resulting conformational map as a probability density in the $(R_g/\ell_c, R_e/\ell_c)$ plane for different temperatures, where $\ell_c$ is the worm contour length and the color code indicates the joint probability density. In free environments, no distinct conformational states emerge, and the distributions remain broad, nearly uniform, and largely temperature-independent.

\subsection*{Active filaments in confinement: Translocation experiments}

In our translocation experiment, we place a single worm (contour length $\ell_c = [15$–$35] \pm 2$~mm and width $w = 0.5 \pm 0.1$~mm) into one of two circular cavities (diameter $d = 8$~mm) connected by a narrow bridge ($w_b = 2$~mm, $L_b = 8$~cm). The channel dimensions exceed the worm’s maximum width and length, allowing the filament to bend, curl, and eventually turn while providing strong confinement conditions. The setup is immersed in a thermostated water bath, enabling precise temperature control and thereby tuning the worm’s activity~\cite{Deblais2020a,Deblais2020b,Sinaasappe2025}. 

The worm motion is recorded for 60~min using a top-view camera equipped with a macro lens (Fig.~S1), with videos typically acquired at a few frames per second to capture the long-timescale dynamics.

We first investigate the locomotion of worms at ambient temperature ($T = 20^{\circ}$C). All conformational and dynamical characteristics reported at this temperature are obtained by averaging trajectories from at least 30 individual worms, each collected in independent experiments. Figure~\ref{fig:Worm_Dynamics_in_Channel}\textit{A} shows a typical sequence of snapshots that illustrate a worm navigating the two-cavity system. Over time, the worm intermittently hops between the two circular cavities, escaping through the aperture and transiting through the constriction (Movie~S1). A representative trajectory of the center-of-mass position along the main axis of the system, $x_{cm}$, is shown in Figure~\ref{fig:Worm_Dynamics_in_Channel}\textit{C}. Notably, the worm does not always cross the bridge in a straight path; instead, it occasionally folds and loops back within the channel before completing the transition. As mentioned, this behavior is mitigated by our choice of bridge width, which forces the worm to actively reptate through the channel~\cite{Sinaasappe2025}.

To check this behavior in simulations, we transfer the tangentially driven polymer model, calibrated as in Fig.~\ref{fig:FreeEnvironement}\textit{A}, under confinement: simulations reproduce key features observed in \textit{T.~tubifex} worms, including partial translocations where the filament folds and reverses within the bridge (Fig.~\ref{fig:Worm_Dynamics_in_Channel}\textit{A},\textit{B},\textit{C} and Movie~S2). 

To quantify the comparison, we compute the probability density of the center-of-mass position along the main axis of the system for worms of intermediate contour length, $\ell_c = 20 \pm 5$~mm (corresponding, in the model, to polymers with contour length $\ell^*_c= 40 \pm 10~\sigma$). The resulting distributions (Fig.~\ref{fig:Worm_Dynamics_in_Channel}\textit{D}) show excellent agreement between experiments (left) and simulations (right). Both living worms and tangentially driven active polymers similarly sample the two cavities, within a comparable time window; the asymmetry in the distribution reflects a statistical memory effect arising from the initial placement of the worm in one of the two chambers. In addition, both distributions indicate that worms and simulated polymers frequently enter the bridge and partially translocate before returning to the chamber where they were initially placed.

A distinctive feature of active filaments is their ability to adjust their conformation in response to varying confinement or crowding~\cite{Kudrolli2019,Fazelzadeh2023,MartinRoca2024,Sinaasappe2025}. To gain insight into their hopping dynamics, we examine the conformations adopted by both worms and active polymers in our two-state system, shown in the $(R_g/\ell_c, R_e/\ell_c)$ plane in Fig.~\ref{fig:Worm_Dynamics_in_Channel}\textit{E}. 

Both experiments and simulations reveal two dominant states: compact conformations, with low values of $R_g$ and $R_e$, and extended ones with high values of both quantities. These correspond to shapes typically observed in the cavities and in the bridge, respectively, as further illustrated in the spatially-resolved representation in Fig.~S3 and S4.
A difference emerges in Fig.~\ref{fig:Worm_Dynamics_in_Channel}\textit{E}: a population of conformations, observed with high-probability, emerges in simulations. This signal is imputable to hairpin-like shapes, similar to the ones observed in DNA \cite{svoboda2006hairpin,bikard2010}. These are likely underrepresented in experiments due to our limitations in tracking complex shapes, particularly when the living worm's shape overlap on itself. More interestingly, the comparison reveals a consistent strategy for translocation in confined geometries: Active filaments alternate between compact shapes to remain in a cavity or pore and extended ones to cross the bridge.

\begin{figure}
     \includegraphics[width=1\columnwidth]{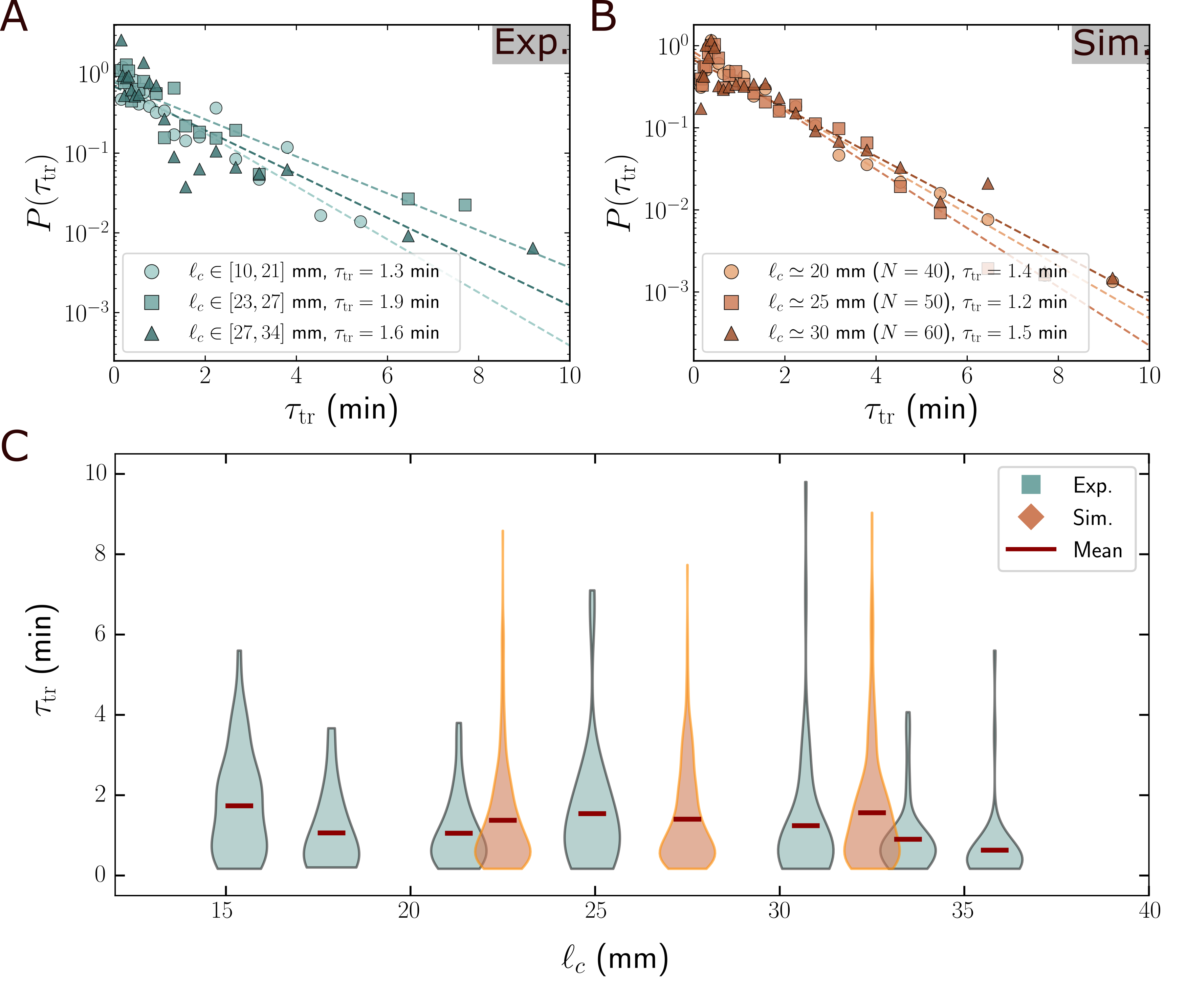}
\caption{\textbf{Effect of the contour length on the trapping time.}
(\textit{A})~Distribution of trapping times $P(\tau_{\mathrm{tr}})$ measured experimentally for worms grouped by contour length into three classes.
(\textit{B})~Distribution of trapping times $P(\tau_{\mathrm{tr}})$ obtained in simulations for three polymer lengths $N=$40, 50, and 60 (with corresponding contour lengths $\ell_c \simeq 20$, 25, and 30~mm). Dashed lines indicate exponential fits.
(\textit{C})~Comparison of trapping times as a function of the worm contour length $\ell_c$, shown as violin plots for both experimental (teal, $T=20\,^{\circ}$C) and simulation (orange, $k_BT^*/u_E=0.2$) data. Each violin shows the distribution of times collected over the full duration of the experiment and the corresponding simulation. Horizontal red bars indicate the mean values.}
	\label{fig:Tau_ContourLength}
\end{figure}

We further examine the dynamics by analyzing the translocation time, defined as the duration required for the center of mass to move from one cavity to the other. This process can be decomposed into two contributions: the trapping time $\tau_{\mathrm{tr}}$, corresponding to the average time spent inside either cavity, and the channel-crossing time $\tau_{\mathrm{ch}}$, associated with transport through the connecting bridge. Since the bridge width $w_b$ is comparable to the filament width, crossings occur predominantly in extended conformations. We therefore focus primarily on the trapping time $\tau_{\mathrm{tr}}$, which dominates the translocation dynamics.

Specifically, we define $\tau_{\mathrm{tr}}$ as the average duration during which the filament’s center of mass, $x_{cm}$, remains within a cavity for a time longer than the characteristic fluctuation timescale of the worm: 
This criterion excludes short-lived excursions into the bridge and ensures that only genuine partial or full translocation events are considered. Indeed, as discussed above, the filament can fold within the channel and re-enter the original cavity, making unsuccessful crossing attempts fairly frequent.

Interestingly, the distribution of trapping times, both in experiments and simulations, follows an exponential law (Fig.~\ref{fig:Tau_ContourLength}\textit{A, B}), in agreement with previous studies on other living worms (\textit{L.~variegatus})~\cite{biswas2023escape} and with an Arrhenius-like process observed for bacteria and active filaments escaping a pore \cite{bhattacharjee2019,bhattacharjee2019b,kurzthaler2021}. In this picture, the pore acts as an entropic trap, with the polymer's activity playing a role analogous to thermal fluctuations.
This suggests that both the conformation and activity of the filament govern its trapping dynamics and motivates a closer examination of how these two key parameters, contour length and activity, affect filament dynamics under confinement.

We thus first compute the trapping times for worms with different contour lengths $\ell_c$. As shown in Figure~\ref{fig:Tau_ContourLength}, the contour length does not significantly affect the probability distribution in either experiments (Fig.~\ref{fig:Tau_ContourLength}\textit{A}) or simulations (Fig.~\ref{fig:Tau_ContourLength}\textit{B}) for characteristic values of $\ell_c \simeq$ 20, 25, and 30~mm. This is further confirmed in Fig.~\ref{fig:Tau_ContourLength}\textit{C}, where the average trapping time remains largely constant across the full range of $\ell_c$ probed in our experiments and simulations (see also Fig.~S5).
This starkly contrasts with thermally driven polymers, whose mobility decreases with increasing contour length, as described by the Rouse and Zimm models \cite{RubinsteinColby2003}. This result instead aligns with a key property of tangentially active filaments: their mobility is independent of contour length \cite{Bianco2018}. It further supports the hypothesis that worms can be considered a tangentially driven system and highlights the advantage of tangential propulsion for filamentous systems navigating complex environments.

\begin{figure}
    \includegraphics[width=\columnwidth]{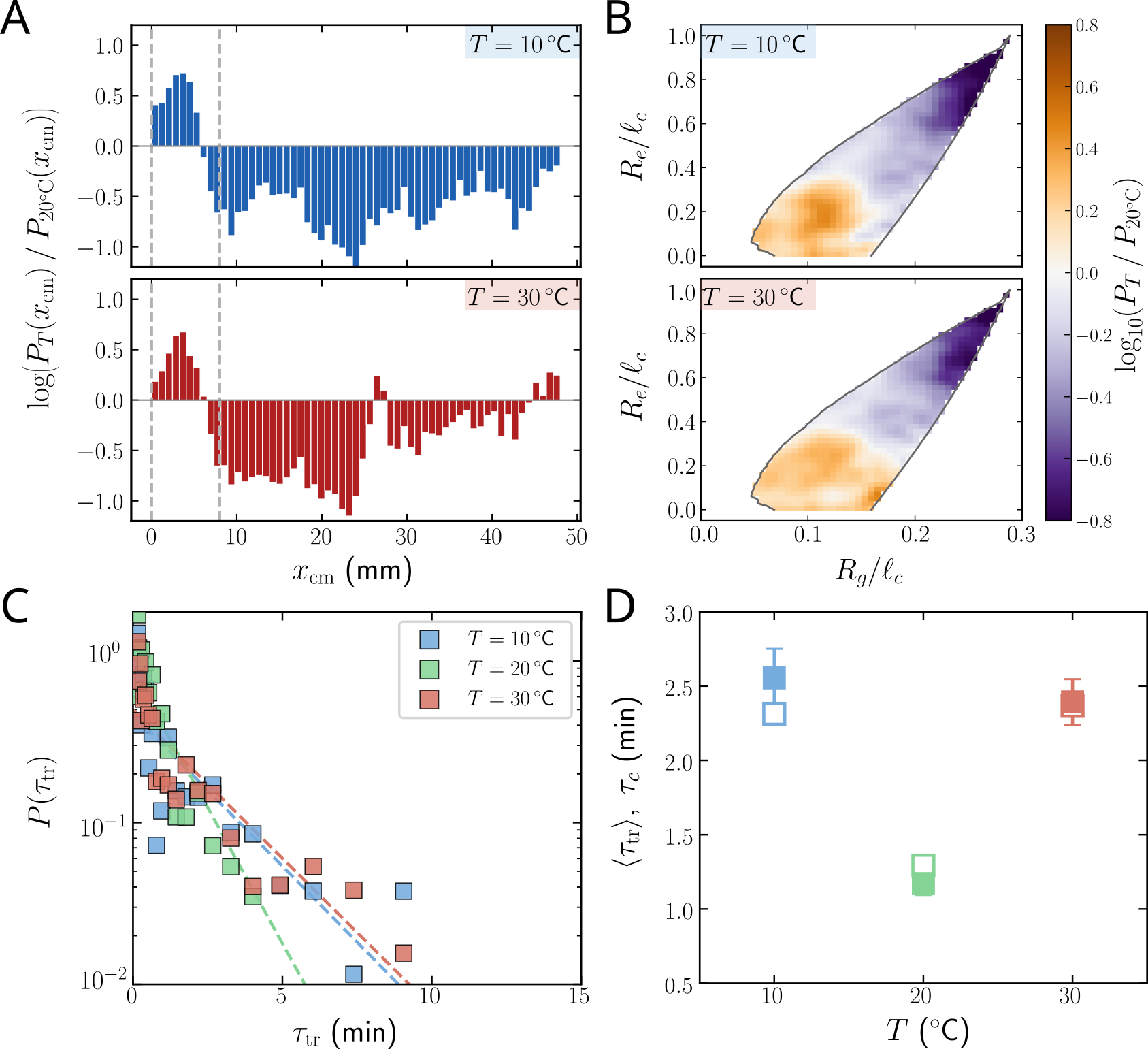}
\caption{\textbf{Effect of activity on worm translocation.}
(\textit{A})~Logarithm of the probability density of the center-of-mass position $P(x_{cm})$ for low ($T = 10^{\circ}$C) and high ($T = 30^{\circ}$C) activity levels, relative to the intermediate case ($T = 20^{\circ}$C). Positive (negative) values indicate an increased (reduced) likelihood. Dashed lines mark the dead end ($x = 0$) and entrance ($x = 8$~mm) of the cavity, in which the worms have been initially placed.
(\textit{B})~Logarithm of the relative probability density of worm conformations in the ($R_g/\ell_c$, $R_e/\ell_c$) plane, normalized by the reference distribution at $T = 20^{\circ}$C shown in Fig.~\ref{fig:Worm_Dynamics_in_Channel}(\textit{E}).
(\textit{C})~Trapping-time distributions for the three temperatures, showing exponential decay with characteristic time $\tau_c$.
(\textit{D})~$\langle \tau_\mathrm{tr} \rangle$ (filled symbols) and $\tau_c$ (open symbols) as a function of $T$ exhibit a minimum at intermediate activity ($T = 20^{\circ}$C), indicating optimal escape from the cavities.}
\label{fig:Heatmap_Effect_of_T}
\end{figure}

We finally examine how activity, tuned via the ambient water temperature~\cite{Deblais2020b} [see SI~\cite{Sup}], influences worm translocation. We focus on three activity levels: low ($T = 10^{\circ}$C), intermediate ($T = 20^{\circ}$C), and high ($T = 30^{\circ}$C). 
As shown in Fig.~\ref{fig:Heatmap_Effect_of_T}\textit{A}, the spatial distribution of the worm’s center of mass, $x_{cm}$, is strongly modulated by temperature. The representation highlights differences between the probability densities of $x_{cm}$ at $T = 10$ and $30^{\circ}$C and the reference at $T = 20^{\circ}$C (Fig.~\ref{fig:Worm_Dynamics_in_Channel}\textit{D}): positive/negative
values indicate higher/lower distributions, compared to the reference. 

At low and high temperatures, worms remain confined to the initial cavity, while intermediate activity allows full exploration of the two-state geometry (Fig.~\ref{fig:Worm_Dynamics_in_Channel}\textit{D}). This is further confirmed by comparing the probability distributions in the ($R_g/\ell_c, R_e/\ell_c$) plane in Fig.~\ref{fig:Heatmap_Effect_of_T}\textit{B}, where compact configurations are more probable and elongated ones less represented at low and high $T$ relative to the reference ($T = 20^{\circ}$C).

This difference affects the trapping-time statistics, shown in Figs.~\ref{fig:Heatmap_Effect_of_T}\textit{C},\textit{D} as a function of temperature. The distributions follow $P(\tau_{\mathrm{tr}}) \propto \exp({-\tau_{\mathrm{tr}}/\tau_c})$, with $\tau_c \simeq \langle \tau_\mathrm{tr} \rangle$; the very good agreement between these quantities supports the exponential fit [Fig.~\ref{fig:Heatmap_Effect_of_T}\textit{C}]. Importantly, $\tau_c$ depends on activity with a non-monotonic response, in contrast to passive systems where higher temperature simply lowers energy barriers and trapping~\cite{muthukumar1999polymer}.

\subsection*{Conformational entropy rules escape dynamics}

\begin{figure*}
    \centering
    \includegraphics[width=0.9\linewidth]{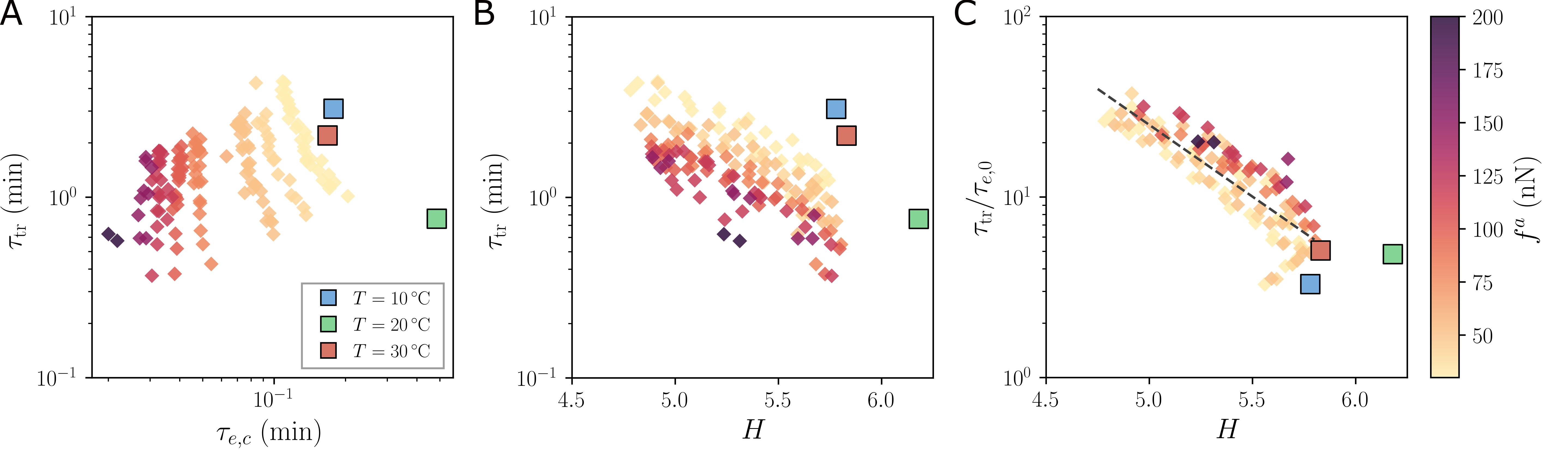}
    \caption{\textbf{Effect of conformational Shannon entropy on trapping time.}
    (\textit{A})~Trapping time versus the decorrelation time in the cavity, for worms (squares) and simulations (diamonds). Simulations span active polymers with $0.3 <\kappa/u_E<1$, $0.1<k_BT^*/u_E<0.3$, and $0.3<f^a \sigma/u_E<2$ (color scale).
    (\textit{B})~Trapping time as a function of the Shannon entropy computed from the conformational maps in the cavity. (\textit{C})~Trapping time, rescaled by the free-space reorientational decorrelation time $\tau_{e,0}$, as a function of the Shannon entropy, following $\tau_{\mathrm{tr}} \propto \tau_{e,0}\,e^{-H}$, highlighted by the black dashed line.}
    \label{fig:entropy}
\end{figure*}

To identify the key parameters controlling escape from the cavity, we systematically vary the active force $f^a$, the effective temperature $T^*$, and the bending stiffness $\kappa$ of the tangentially driven polymer. In Fig.~\ref{fig:entropy}\textit{A}, we report the trapping time as a function of the characteristic decorrelation time of shape fluctuations in the cavity, $\tau_{e,c}$~\cite{sinaasappel2026}. Since $\tau_{e,c}$ depends on all model parameters ($T^*$, $\kappa$, and $f^a$), we highlight here the role of activity through the color scale (for the dependence with the stiffness, see Fig.~S6). Weakly active polymers (light colors) exhibit a strong dependence of the trapping time on $\tau_{e,c}$, whereas highly active polymers are only weakly affected. Both quantities decrease markedly with increasing activity $f^a$, reflecting faster conformational rearrangements and more efficient escape dynamics. These observations demonstrate that activity, stiffness, and thermal fluctuations (here mimicking biological noise) jointly control both the conformational dynamics and the translocation process under confinement.

To further connect conformations and dynamics, we reduce the two-dimensional conformational space $(R_g/\ell_c, R_e/\ell_c)$ under confinement to a single scalar quantity, the Shannon entropy $H$~\cite{lin1991}, which quantifies the diversity of explored conformations. $H$ is computed from the joint probability distribution of the conformational map as
\begin{equation}
    H = -\sum_i p_i \ln p_i,
\end{equation}
where $p_i$ is the probability of finding the filament in bin $i$ of the discretized $(R_g/\ell_c, R_e/\ell_c)$ plane. By construction, $H$ is maximal when all accessible conformations are equally likely, \emph{i.e.}, for a broad and uniform distribution, and minimal when the filament is restricted to a few conformations. This coarse-grained description captures, in a single observable, the extent of conformational exploration and enables a direct comparison between simulations (diamonds) and experiments on living worms (squares).

The results are shown in Fig.~\ref{fig:entropy}\textit{B} for the same range of activity as in Fig.~\ref{fig:entropy}\textit{A} (see color code). Since $H$ remains only weakly dependent on the active force $f^a$, this representation isolates more clearly the effect of temperature and stiffness (see Fig.~S8) on the trapping dynamics. In particular, larger conformational entropy systematically correlates with shorter trapping times, indicating that broader exploration of conformational space facilitates escape from the cavity.

Notably, living worms at ambient temperature occupy a region of significantly higher entropy that is not accessible to the minimal tangentially driven polymer model within the range of $f^a$, $T^*$, and $\kappa$ explored here. This further suggests that conformational entropy is a particularly relevant quantity for characterizing transport in living filamentary systems, whose dynamics likely involve additional internal degrees of freedom beyond homogeneous tangential propulsion.

Remarkably, when the trapping time is normalized by the decorrelation timescale in free space, $\tau_{e,0}$ (Fig.~\ref{fig:entropy}\textit{C}), all data collapse onto a single master curve following
\[
\tau_{\mathrm{tr}} \propto \tau_{e,0} \, e^{-H}.
\]
This collapse demonstrates that escape is governed by two coupled ingredients: the intrinsic timescale over which the filament reconfigures, and the extent to which it uniformly explores its conformational space. In this framework, $H$ characterizes the shape of an effective conformational landscape. When conformations are sampled more uniformly, the filament is less likely to remain trapped in a restricted subset of shapes and can more efficiently access configurations favorable for escape. As a result, broader conformational exploration leads to shorter trapping times.

This approach further provides a predictive description of active polymer translocation: the escape time can be inferred directly from macroscopic conformational observables such as $R_g$ and $R_e$, through the entropy $H$, without detailed knowledge of the microscopic dynamics. It also connects naturally to the classical picture of entropic trapping in passive polymers, where confinement generates an effective free-energy barrier set by geometry, such as the pore size~\cite{bhattacharjee2019b,kurzthaler2021,biswas2023escape}. In the present case, activity does not primarily modify the barrier itself, but instead controls how efficiently the filament explores the underlying conformational landscape and accesses escape-compatible configurations.

Interestingly, within this framework, living \textit{T.~tubifex} worms achieve optimal translocation at ambient temperature ($T = 20^\circ$C), corresponding to maximal conformational entropy under confinement.

\subsection*{Optimal escape emerges from the interplay of rotation and translation}

\begin{figure*} 
\centering
\includegraphics[width=0.9\linewidth]{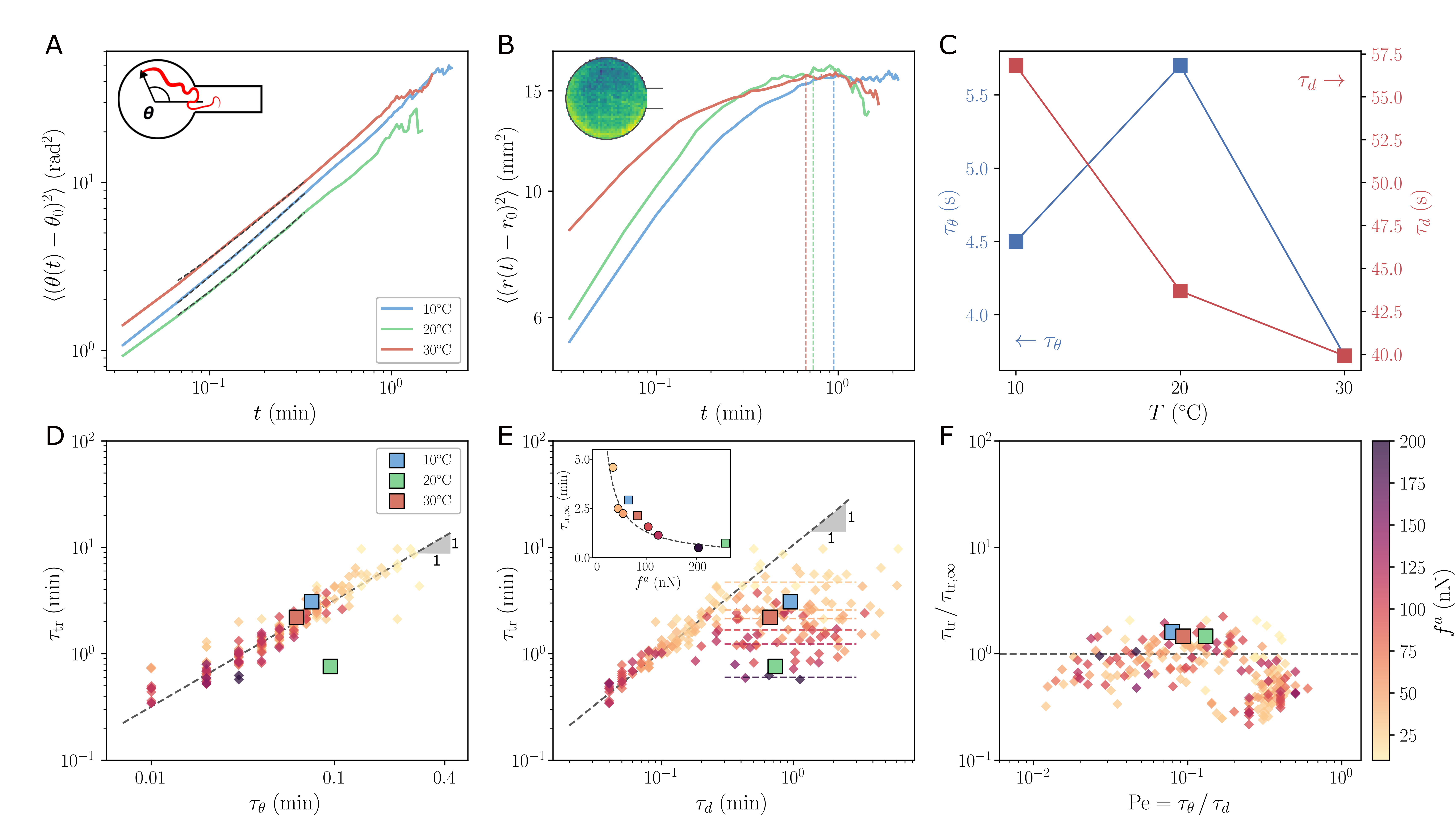}
\caption{\textbf{Diffusion of living and active polymers in confinement and criterion for optimal translocation.} 
(\textit{A},\textit{B})~Rotational and translational mean square displacements of the worm's head within the cavity. Insets: definition of the angular coordinate of the head position inside the cavity (\textit{A}) and probability density function of the worm head position within the cavity (\textit{B}). 
(\textit{C})~Rotational and translational timescales, $\tau_{\theta}$ and $\tau_{\mathrm{d}}$, extracted from (\textit{A}) and (\textit{B}), respectively, as a function of temperature. 
(\textit{D}),(\textit{E})~Trapping time as a function of the rotational timescale and the translational diffusion. The inset in (\textit{E}) shows the dependence of the plateau trapping time $\tau_{\mathrm{tr}, \infty}$, obtained in the limit of long translational times $\tau_{\mathrm{d}}$, on the active force $f^a$. 
(\textit{F})~Characteristic trapping time $\tau_{\mathrm{tr}}$, normalized by the plateau value, plotted as a function of the P\'eclet-like number, $\mathrm{Pe} = \tau_{\theta}/\tau_{\mathrm{d}}$.} 
\label{fig:fig6_trap_criteria} 
\end{figure*}

While this entropy-based description successfully captures the relation between conformational exploration and escape dynamics, it does not yet identify the physical mechanism underlying this optimality. To address this question, we now focus on the dynamics of the worm’s head inside the cavity. For tangentially driven active polymers, as well as for living worms, head dynamics has been shown to control key aspects of transport under confinement~\cite{biswas2023escape,kapadia2026}.

We characterize both translational motion, via the mean square displacement (MSD), and rotational motion [Fig.~\ref{fig:fig6_trap_criteria}\textit{A},\textit{B}] inside the cavity. For the latter, we adopt a reference frame centered on the centre of the cavity, in which the head position defines an angle $\theta$ with respect to the main axis of the channel (Fig.~\ref{fig:fig6_trap_criteria}\textit{A}, inset). The angular mean square displacement, $\langle (\theta(t) - \theta_0)^2 \rangle$ (rMSD), is computed by unwrapping $\theta(t)$ along the infinite semi-axis $(0, +\infty)$ for trajectories where the worm’s head remains within a single cavity. The rMSD thus quantifies the angular exploration of the head prior to locating the exit.

Unlike \emph{L.~variegatus}, which remains predominantly near the cavity walls~\cite{biswas2023escape}, the head of \emph{T.~tubifex} explores the cavity more uniformly (Fig.~\ref{fig:fig6_trap_criteria}\textit{B}, inset). The rotational dynamics exhibit a nonmonotonic trend: at $T = 20^\circ$C, worms show reduced angular exploration, indicating less time spent reorienting within the cavity (Fig.~\ref{fig:fig6_trap_criteria}\textit{A}). In contrast, translational mobility increases monotonically with temperature (Fig.~\ref{fig:fig6_trap_criteria}\textit{B}), consistent with enhanced activity~\cite{Deblais2020b,Sinaasappe2025}.

To quantify these two processes, we extract two characteristic timescales from the MSD and rMSD (Fig.~\ref{fig:fig6_trap_criteria}\textit{C}): the rotational timescale $\tau_{\theta}$, set by the angular diffusion coefficient obtained from a fit of the rMSD (Fig.~\ref{fig:fig6_trap_criteria}\textit{A}), and the translational timescale $\tau_{\mathrm{d}}$, defined as the time at which the head's MSD saturates due to the finite cavity size (Fig.~\ref{fig:fig6_trap_criteria}\textit{B}).

We further employ simulations to understand how these quantities control escape dynamics (Fig.~\ref{fig:fig6_trap_criteria}\textit{D},\textit{E}). The trapping time displays a linear dependence on the rotational timescale, confirming that slow reorientation limits escape from the cavity. In contrast, its dependence on translational dynamics is markedly nontrivial (Fig.~\ref{fig:fig6_trap_criteria}\textit{E}). At high translational diffusion (i.e., small $\tau_{\mathrm{d}}$), the trapping time is nearly independent of the active force and increases with increasing $\tau_{\mathrm{d}}$. In this regime, the filaments are typically rather stiff (see Fig.~S7), closely follow the cavity boundaries, and they escape in a quasi-deterministic manner; consequently, their escape dynamics correlate directly with translational mobility. 

At small translational diffusion (i.e., high $\tau_{\mathrm{d}}$), a second regime emerges in which the trapping time saturates toward a plateau value, denoted $\tau_{\mathrm{tr},\infty}$. Increasing activity shifts the crossover to smaller values of $\tau_{\mathrm{d}}$. In this regime, escape is dominated by stochastic exploration: the filament locates the exit primarily by chance, such that faster translational motion does not necessarily enhance escape efficiency. Living worms populate this second regime, consistent both with the stochastic exploration observed experimentally (inset of Fig.~\ref{fig:fig6_trap_criteria}\textit{B}) and with the highly flexible nature of \textit{T.~tubifex}.

This interpretation suggests an alternative physical picture for the optimal translocation observed experimentally: Under optimal environmental conditions, worms effectively propel more efficiently. Indeed, plotting the plateau trapping time $\tau_{\mathrm{tr},\infty}$ extracted from simulations as a function of activity $f^a$ [inset of Fig.~\ref{fig:fig6_trap_criteria}\textit{E}] reveals a power-law scaling,
\[
\tau_\mathrm{tr}^\infty \propto \frac{1}{f^a}.
\]
Using this relation, we can assign an effective active force to worms at different temperatures (see SI). Within this framework, worms at $T = 20^\circ$C behave as confined active filaments with a larger effective propulsion force than worms at lower or higher temperatures.

Finally, we combine these observations in Fig.~\ref{fig:fig6_trap_criteria}\textit{F}, where we plot the trapping time, normalized by the plateau value $\tau_{\mathrm{tr},\infty}$, as a function of a P\'eclet-like number defined as
\[
\mathrm{Pe} = \frac{\tau_{\theta}}{\tau_{\mathrm{d}}},
\]
which compares the timescale of rotational reorientation to that of translational exploration within the cavity. Experimental values are normalized using the predicted plateau trapping time extracted from the simulation scaling law (see SI).

In this representation, both experimental and simulation data collapse onto a weakly varying master curve. This indicates that the trapping dynamics are primarily set by the activity-dependent plateau time $\tau_{\mathrm{tr},\infty}$, regardless of the relative balance between rotational reorientation and translational exploration. The collapse further reveals a common escape mechanism for living worms and tangentially driven active filaments under confinement. 

Notably, ``deterministic'' filaments are characterized by high values of $\mathrm{Pe}$ and lie below unity in this representation. Experimental data, by construction, cluster around $\mathrm{Pe} \sim 0.1$ and correspond to intermediate activity levels, which peak at the optimal temperature. 

As such, the worm's optimality emerges from the competition between propulsion and orientational dynamics. At low activity, the worm lacks sufficient drive to reach the exit; at high activity, excessive orientational persistence hinders efficient reorientation, leading to self-trapping. 

These observations further clarify the interplay between translational exploration and rotational reorientation governing filament escape, and suggest a possible design principle for sorting active filaments. By tuning cavity and aperture sizes, one could select filaments according to their activity, potentially via local control of temperature or other environmental parameters.

In conclusion, our study reveals the distinct translocation dynamics of active polymer-like \textit{T.~tubifex} worms and polar, semi-flexible tangentially driven polymers in confined geometries, demonstrating that their transport properties differ fundamentally from those of passive polymers, for which increasing temperature enhances diffusion by lowering trapping times. In contrast, we find a non-monotonic dependence of translocation efficiency on activity: an intermediate activity level maximizes exploration, while both lower and higher activities hinder translocation. This behavior arises from the interplay between directed motion and rotational diffusion, which together control the efficiency of cavity escape and bridge crossing.  

By systematically comparing experiments with simulations of tangentially driven active polymers, we show that homogeneously distributed active forces capture the essential features of the observed dynamics. Moreover, the negligible influence of contour length on translocation highlights activity as the primary control parameter. This validates our approach and supports the use of tangentially driven polymers as a minimal framework to describe living worm dynamics and, more broadly, active (polar) polymer transport in confinement.  

The comparison also supports the introduction of Shannon entropy as a coarse-grained description that captures, in a single quantity, the extent to which the filament samples compact and extended configurations, and thus its ability to find escape pathways. Indeed, the entropy is maximized by worms at intermediate temperature, consistent with what is expected for a species adapted to temperate climates, suggesting efficient cycling through different conformational states.  

Beyond this specific system, our results point to general principles governing filamentous and polymer transport in confined active matter. The existence of an optimal activity level suggests a strategy to enhance translocation efficiency, which could inform the design of biomimetic active filaments \cite{xi2024,Al-Izzi2026}. Future work could extend these ideas to more complex environments, such as disordered landscapes or structured microfluidic networks, to further elucidate how activity-driven transport is shaped by geometry, boundary interactions, and external fields such as flow or electric forcing.

\section*{Methods}
\subsection*{Living Worms}
All batches of \textit{T.~tubifex} worms analyzed in this work were purchased from the provider Aquadip (\url{https://www.aquadip.nl/}) and ordered in a prepacked configuration, where the worms were at adult size. 
The worms were maintained in an aquarium at room temperature, constantly under filtered flow, with water consisting of demineralized water mixed with salt solutions optimized for their needs. 
The salt solution consist of a mixture of: 50~g/L of \ce{NaHCO3}; 10~g/L of \ce{KHCO3}; 100~g/L of \ce{CaCl2}; 90~g/L of \ce{MgSO4}.  
Worms were fed weekly with standard goldfish food, and the water was refreshed once per week or more frequently if needed. After undergoing a one hour habituation period, worms were transferred in the experimental setup.

\subsection*{Numerical model}
The worm body is modeled as a semi-flexible, self-avoiding, bead-spring linear polymer in 3 dimensions. 
The polymer is made of $N=\ell_c/w$ monomers where $\ell_c$ is the worm's length and $w$ is its average thickness, mapping the monomer size to $w$.
We employ a modified Kremer-Grest polymer model~\cite{kremer1990dynamics}, introducing polymer rigidity with a bending potential (see SI for more details).
Since worms sediment, we include the effect of gravity through the addition of a constant acceleration $g$ along the $z$ direction. This requires to introduce a surface at $z=0$, modeled as a perfectly flat, purely repulsive wall. 
As a minimal model for the worms' peristaltic propulsion, we employ the tangential propulsion model~\cite{Bianco2018}: each active monomer is subject to an active force that is parallel to the (normalized) backbone tangent vector; the end monomers are always passive. 
Finally, because of the macroscopic nature of the system, thermal fluctuations are negligible. 
However, worms constantly move around and explore their surroundings, producing a random motion; we approximate this motion through a fictitious thermal (that is, uncorrelated) noise and, since the worms are immersed in water, we employ a Langevin thermostat, disregarding hydrodynamic interactions, because of the presence of the surface.

\begin{acknowledgments}
We thank the Technology Center of the University of Amsterdam for technical support. C.V. acknowledges funding from IHRC22/00002 and Proyecto PID2022-140407NB-C21 by MCIN/AEI/10.13039/501100011033 and FEDER, UE. M.V. acknowledges support from the São Paulo Research Foundation (FAPESP) under grant \#2025/26193-7. M.V. and E.L. acknowledge the CINECA-INFN agreement, providing access to
resource on Leonardo at the CINECA computational
infrastructure; CloudVeneto is also acknowledged for the use of computing
and storage facilities.
\end{acknowledgments}

\clearpage

\onecolumngrid 

\setcounter{secnumdepth}{3}
\setcounter{section}{0}
\setcounter{subsection}{0}
\setcounter{equation}{0}
\setcounter{figure}{0}
\setcounter{table}{0}

\renewcommand{\thesection}{S\arabic{section}}

\renewcommand{\theequation}{S\arabic{equation}}
\renewcommand{\thefigure}{S\arabic{figure}}
\renewcommand{\thetable}{S\arabic{table}}

\begin{center}
\textbf{\large \vspace*{1.5mm} Optimal Translocation of Living \& Active Filaments in Confinement -- Supplementary Material } \\
\end{center}
\vspace*{5mm}

\section{Experimental details}
\subsection{\textit{Tubifex tubifex} worms}

The living worms used in this study were \textit{Tubifex tubifex}, a living system previously investigated in recent studies \cite{Deblais2020a,Deblais2020b,Heeremans2022,Deblais2023,Sinaasappe2025}. All batches of \textit{T.~tubifex} worms analyzed in this work were purchased from the provider Aquadip (\url{https://www.aquadip.nl/}) and ordered in a prepacked configuration, where the worms were at adult size. The worms were maintained in an aquarium at room temperature, constantly under filtered flow, with water consisting of demineralized water mixed with salt solutions optimized for their needs. 
The salt solution consist of a mixture of: 50~g/L of \ce{NaHCO3};  10~g/L of \ce{KHCO3}; 100~g/L of \ce{CaCl2}; 90~g/L of \ce{MgSO4}.  
Worms were fed weekly with standard goldfish food, and the water was refreshed once per week or more frequently if needed.

\subsection{Experimental set-up and Image acquisition}

\begin{figure}[h]
\centering
\includegraphics[width=1\textwidth]{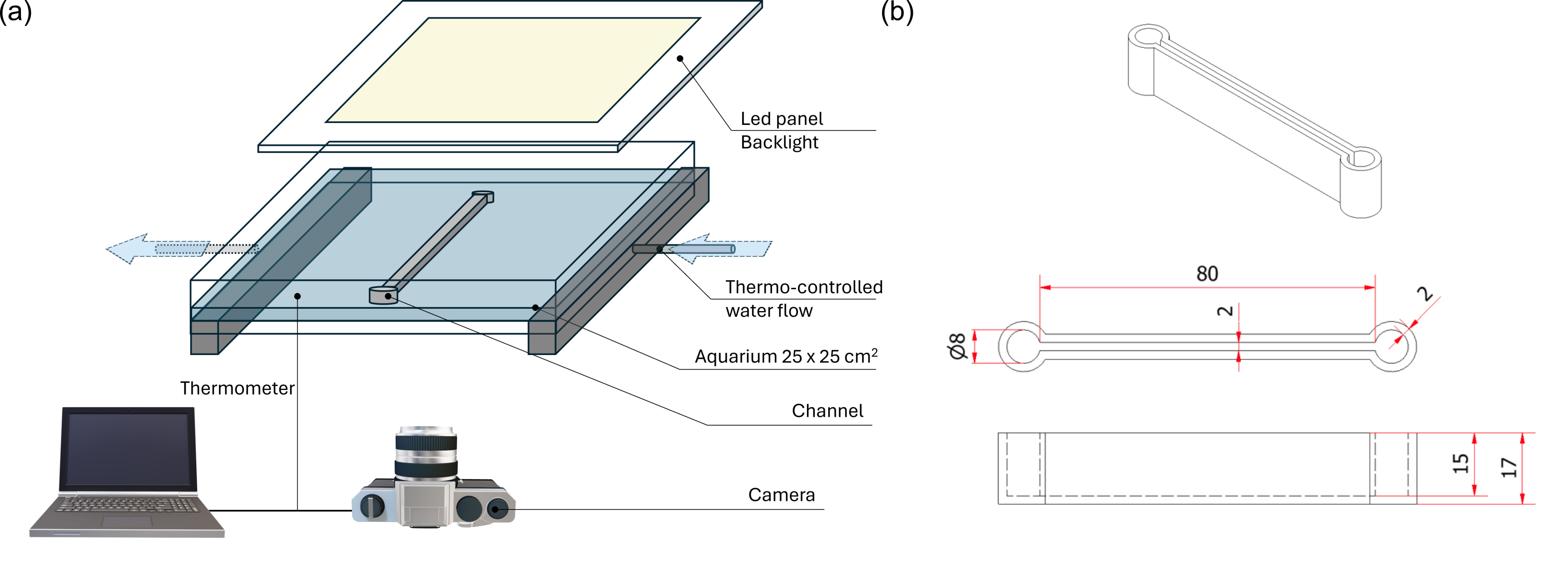}
\caption{\textbf{Experimental setup.}
(a)~Schematic of the experimental setup used to characterize the dynamics and conformations of \textit{T.~tubifex} worms under confinement (not to scale). Images are acquired from below using a camera positioned beneath the aquarium, minimizing visual obstruction by the channel walls and facilitating worm tracking. The aquarium is filled with water maintained at a controlled temperature $T$.
(b)~3D-printed confinement geometry used in the translocation experiments. The channel consists of two circular chambers connected by a narrow bridge; dimensions are indicated in mm. The device is fabricated from a transparent material to enable imaging with transmitted light.
}
\label{SupFig:Experimental_Setup}
\end{figure}

For each measurement, worms were transferred and separated using a pipette. Due to natural size variations associated with age, worms of different contour length $\ell$ were selected (see main text). An individual worm was then placed in the 3D-printed channel within a temperature-controlled reservoir filled with aquarium water. A photograph and schematic of the whole experimental setup, including the 3D-printed channel is shown in Fig.~\ref{SupFig:Experimental_Setup}.  

The water temperature was regulated using a thermo-controlled bath that circulated hot or cold water through the bottom of the reservoir. The resulting temperature was monitored with a digital probe thermometer, and measurements were conducted at low (10$^\circ$C), intermediate (room temperature, 20$^\circ$C), and high (30$^\circ$C) temperatures.  

Both the two-state system and reservoir have transparent bottoms, allowing a camera (Nikon D5300) to be positioned below the reservoir while a white LED panel provided back-lighting from above. The camera captured sequences of images as the worm moved within the channel. An example image sequence is shown in Figure~1 of the main text, and a schematic of the setup is presented in Fig.~\ref{SupFig:Experimental_Setup}.

\subsection{Image processing}
\label{sec:img_proc}
\begin{figure}[h]
\centering
\includegraphics[width=0.5\textwidth]{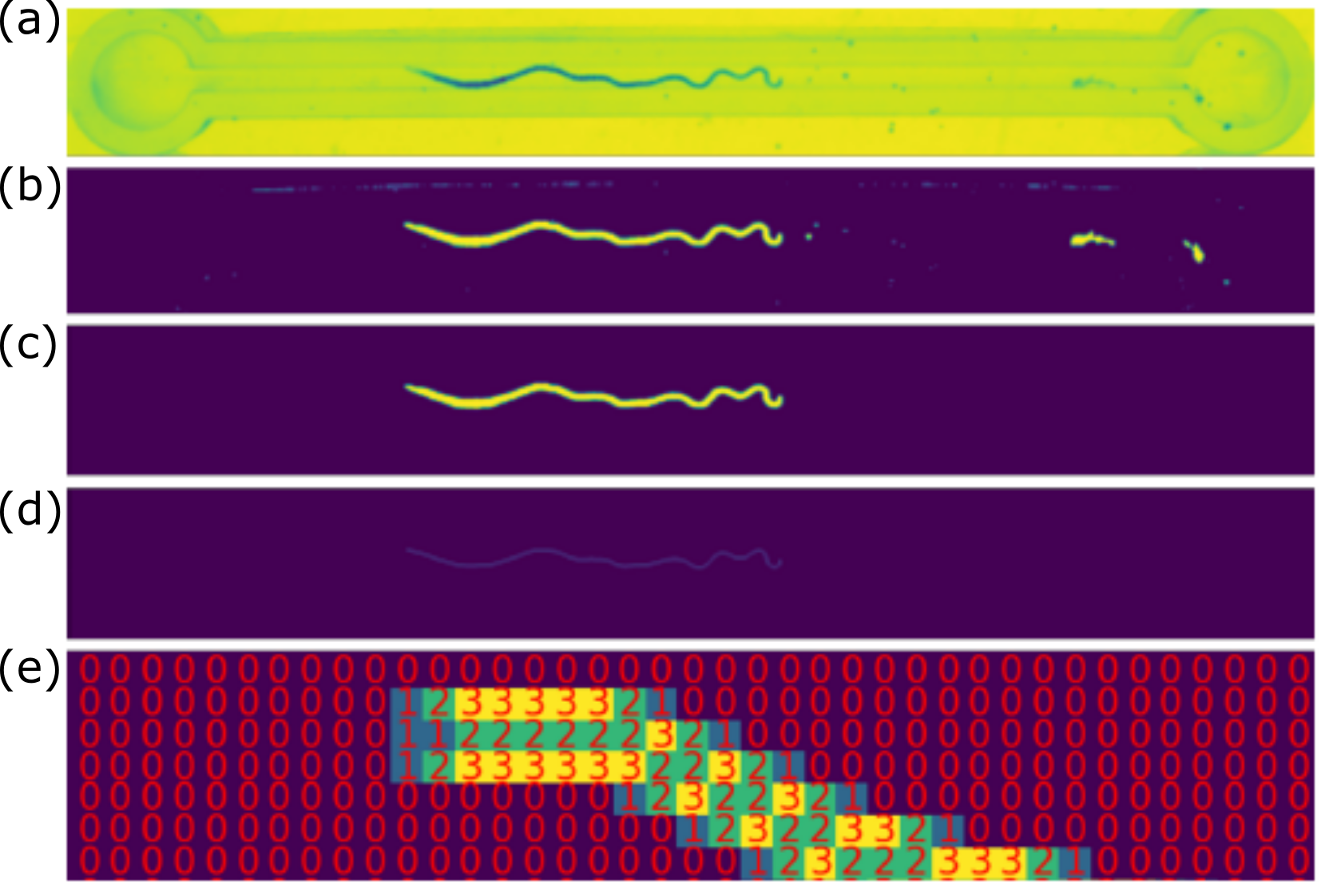}
\caption{\textbf{Image-processing pipeline for living worms detection and tracking.}
Steps of the image-analysis procedure used to extract the worm contour and center-of-mass over time. 
(a)~Representative image of a living \textit{T.~tubifex} worm confined within the channel. 
(b)~Background subtraction and thresholding to obtain a binary image. 
(c)~Identification and selection of connected pixel clusters corresponding to the worm. 
(d)~Skeletonization of the segmented worm shape to extract its centerline. 
(e)~Detection of the worm endpoints based on the local connectivity of skeleton pixels.}
\label{SupFig:Image_Processing}
\end{figure}

The images of the worms were analyzed using a custom Python script. The different steps of the image processing are shown in Fig.~\ref{SupFig:Image_Processing}.  

First, each frame was loaded into the script and cropped to the region containing the channel. In the initial iteration, cropping coordinates were manually defined. A background image was then subtracted to remove as many non-worm pixels as possible, followed by conversion to a binary image. However, background subtraction alone was often insufficient, as small air bubbles or impurities remained. To address this, only the largest group of connected pixels was retained, which consistently corresponded to the worm when images were acquired carefully.  

At this stage, the center-of-mass coordinates $(x,y)$ and the radius of gyration $R_g$ were computed. To determine the worm’s endpoints, the image was skeletonized, reducing the worm to a one-pixel-wide line. A convolution was then applied to detect the number of neighboring pixels for each pixel. The endpoints were identified as pixels belonging to the skeleton with only one neighbor. We define the end-to-end distance $R_e$.  

This process was repeated for every frame in the measurement. In subsequent iterations, manual cropping was replaced by automatic cropping centered around the worm’s previous center-of-mass position, significantly reducing computation time through more efficient processing.  

\clearpage
\section{Model and simulations details}
\subsection{Model}

\paragraph{Bead-spring filament model}
The worm body is modeled as a semi-flexible, self-avoiding, bead-spring linear chain in 3D consisting of a number $N$ of monomers, related to its length $\ell$ via the relation $N=\ell/w$ where $w$ is the worm's average thickness ($w=0.5$ mm); we map the characteristic monomer size ($\sigma=1$) to $w$. Self-avoidance between any pair of monomers is implemented via a truncated and shifted Lennard-Jones (LJ) potential: 
\begin{equation}
V_{\text{LJ}}(r) =
\begin{cases}
  4\epsilon \left[ \left(\frac{\sigma}{r} \right)^{12}- \left(\frac{\sigma}{r}\right)^{6}+\frac{1}{4}\right] & \text{for~}  r < 2^{1/6}\sigma \\
  0 & \text{for~} r \geq 2^{1/6}\sigma
\end{cases}
\end{equation}
where $\sigma=1$ is the characteristic monomer size, $\epsilon=10\,u_E$ with $u_E$ the scale of energy (see below) and $r=|\vec{r}_i - \vec{r}_j|$ is the Euclidean distance between the monomers $i$ and $j$ positioned at $\vec{r}_i$ and $\vec{r}_j$, respectively. 
The Finitely Extensible Nonlinear Elastic (FENE) potential~\cite{kremer1990dynamics}
\begin{equation}
	V_\mathrm{FENE}(r) = -\frac{K r_0^2}{2} \ln \left[ 1 - \left( \frac{r}{r_\mathrm{0}} \right)^2 \right]
\end{equation}
acts between any pair of consecutive monomers along the filament. We set $K= 30\,\epsilon/\sigma^2=300\,u_E/\sigma^2$ and $r_0=1.5\sigma$ in order to avoid strand crossings. Finally, we introduce filament rigidity with the bending potential:
\begin{equation}
\label{Ub}
U_{b} (\theta) = \kappa (1+\cos{\theta}),
\end{equation}
where $\theta$ is the angle formed by three consecutive beads along the backbone and $\kappa$ is the bending energy.

Monomers have unitary mass ($m=1$); we can remap the simulation mass to the experiments via the density of the worm, such that the mass of a single ``bead'' is given by $m^* = \frac{\pi}{4} (\rho-\rho') w^3$, where $\rho= 1.1\,\mathrm{g/cm^3}$ and $\rho'$ is the density of water. The result follows the approximation of the worm as a cylinder of radius $w/2$ and length $\ell=Nw$. Notice that worms have a higher density with respect to water and, being macroscopic, sediment. We implement this key physical feature by introducing the effect of gravity, that is, there is constant acceleration $g$ along the $z$ direction. This requires to introduce a surface at $z=0$, modeled as a perfectly flat, purely repulsive wall that interacts with each monomer via  
\begin{equation}
V_{\text{wall}}(r) =
\begin{cases}
  4\epsilon \left[ \left(\frac{\sigma}{z} \right)^{12}- \left(\frac{\sigma}{z}\right)^{6}+\frac{1}{4}\right] & \text{for~}  z < 2^{1/6}\sigma \\
  0 & \text{for~} z \geq 2^{1/6}\sigma.
\end{cases}
\end{equation}
As a consequence of the action of gravity, the filament moves prevalently on the surface. However, the motion is not restricted in 2D but is rather still three-dimensional, enabling detachment from the surface, a behaviour observed also in experiments. Notably, we map the unit of energy of the model with the work done by gravity on the scale of the monomer size, $u_E=1=mg\sigma$. As such, we map the rescaled units of the model to the real units of the experiment as:
\begin{equation}
    \sigma = 1 \to w = 0.5\,\mathrm{mm}; \qquad m=1 \to m^*= 10^{-5}\,\mathrm{g}; \qquad u_E = 1 \to u_E^* = m^* g w = 5 \cdot 10^{-11}\,\mathrm{J}. 
\end{equation}
We can thus set the unit of time to $\tau=\sqrt{m \sigma^2/u_E}$, which corresponds to $\tau^* = \sqrt{w/g}=0.007\,\mathrm{s}$.

Finally, because of the macroscopic nature of the system, thermal fluctuations are negligible; notice that, indeed, $u_E \gg k_B T$ at room temperature. However, worms constantly move around and explore their surroundings, producing a random motion; we approximate this motion through a fictitious thermal (that is, uncorrelated) noise, characterized by a putative temperature $T$. Due to the fact that the worms are immersed in water, we introduce a Langevin thermostat, with friction coefficient $\gamma=20 u_E \,\tau/\sigma^2$. We disregard hydrodynamic interactions, because of the presence of the surface.\\
A brief note on the value of $\gamma$: it is unfortunately not straightforward to map the friction coefficient in the present context. An approximate calculation reads as follow: the sedimentation velocity, neglecting thermal fluctuations, reads 
$$v_{\mathrm{sed}}=\frac{2}{9}\frac{(\rho-\rho')g(w/2)^2}{\eta},$$ 
where $\eta$ is the viscosity. The typical time scale is given by $\tau_{\mathrm{s}} = {w}/{v_{\mathrm{sed}}}\simeq 4 \cdot 10^{-2}\,\mathrm{s}$ given the viscosity of water at room temperature ($\eta\simeq10^{-3} \mathrm{Pa\cdot s}$). In simulation units, $\tau_\mathrm{s}\simeq 5\, \tau$ and in the simulation, where hydrodynamic interactions are absent, the sedimentation velocity reads $\gamma {\sigma^2}/{\tau_\mathrm{s}}=u_E$, leading to $\gamma\simeq 5 u_E \,\tau/\sigma^2$. Thus, we overestimate the value of $\gamma$. We do this for two reasons: (i) the above argument is based on strong approximations and, thus, the obtained value of $\gamma$ should not be taken as a precise estimate; (ii) the Reynolds number associated with the worms' motion is small ($\mathrm{Re}\simeq10^{-1}-10^{1}$) and, thus, ensuring inertia is negligible is crucial to capture the physics of the system.\\

\paragraph{Activity}
We employ the tangential model, as implemented in Ref.~\cite{Bianco2018}. Briefly, activity is introduced as a tangential self-propulsion: an active monomer $i$ at position $\vec{r}_{i}$ is subject to an active force $\vec{f}^{\mathrm{a}}_{i} = f^{\mathrm{a}} \hat{t}_i$ where $\hat{t}_i = (\vec{r}_{i+1} - \vec{r}_{i-1})/|\vec{r}_{i+1} - \vec{r}_{i-1}|$ is the normalized tangent vector. The end monomers are always passive.\\ 

\paragraph{Confinement}
We implement a confining structure, mimicking the experimental one by placing immobile beads of size $\sigma$ in the $xy$ plane at a distance $r_w=1\,\sigma$ from each other. The two chambers have a diameter $D=16\,\sigma$, the connecting channel (or ``bridge'' as in the main text) has a length $L_b = 160\,\sigma$ and width $w_b = 4\,\sigma$. Notice that there is a small discrepancy between the measurements in simulations and in the experiments, as in simulations we measure distances between the centre of mass of the beads, while in experiments distances are measured from the surface of the wall.    

\subsection{Simulation details}
\label{sec:sim}	
The chains are simulated in bulk conditions (free environment) and under confinement. In all cases, simulations are performed using the open source code LAMMPS\cite{thompson2022lammps}, with in-house modifications to implement the tangential activity. We perform Langevin Dynamics simulations disregarding hydrodynamic interactions. The equations of motion are integrated using the velocity Verlet algorithm, with an elementary time step $\Delta t = 10^{-3} \tau$. The units of mass, length, and energy are set to $m=1$, $\sigma=1$, and $u_E = 1$, respectively.\\
In bulk conditions, we simulate filaments of length $N =40$ with different values of the activity parameter $f_a =$ 0.1-2.0 $u_E/\sigma$, $\kappa =$0.2-0.6 $u_E$ and $k_B T$=0.1-0.3 $u_E$. We simulate $M=$10 independent trajectories for each set of parameters. After reaching a steady state, production runs are performed for $\approx 1 \cdot 10^{8}$ time steps and conformations are sampled at a rate of $10^{5}$ time steps, that is, of the same order of magnitude as the decorrelation time of the end-to-end vector. We also performed short runs with a sampling rate of $10^2-10^3$ time steps to sample short-scale dynamics.

Under confinement, we similarly explore the parameter space. In particular, we simulate filaments of length $N = 40$, $f_a =$ 0.1-2.0 $u_E/\sigma$, $\kappa =$0.4-1.0 $u_E$ and $k_BT$=0.1-0.3 $u_E$.
We simulate $M=$10 independent trajectories for each set of parameters; worms are initially always placed in the same chamber. As in experiments, runs are performed for $5 \cdot 10^{8}$ time steps, corresponding to $\approx 3600$ seconds, and conformations are sampled at a rate of $10^{4}$ time steps.

\subsection{Fitting of $\langle \tau_{\mathrm{tr}}^{\infty} \rangle$ vs $f^a$}

We start from a simulation dataset spanning different values of $f^a$, $T^*$, and $\kappa$. From this dataset, we first select data at fixed $f^a$ and then extract all points belonging to the plateau regime at large $\tau_{\mathrm{d}}$; averaging over these points yields $\langle \tau_{\mathrm{tr}}^{\infty} \rangle$ for each value of $f^a$. The resulting values are shown in the inset of Fig.~6E of the main text.

We assume a power-law dependence,
\begin{equation}
\langle \tau_{\mathrm{tr}}^{\infty} \rangle = A (f^a)^\alpha,
\label{eq:tautr_fit}
\end{equation}
and perform a linear fit of $\ln(\langle \tau_{\mathrm{tr}}^{\infty} \rangle)$ as a function of $\ln(f^a)$. This yields an exponent $\alpha = -0.95 \pm 0.02$, consistent with the value $\alpha = -1$ reported in the main text. A similar scaling has been observed for active Brownian particles~\cite{debnath2021escape}.

To assign an effective active force to the experimental data, one could in principle invert ~\eqref{eq:tautr_fit}. However, a direct inversion of the mean trapping time would then forbid to use Eq.~\eqref{eq:tautr_fit} to estimate $\langle \tau_{\mathrm{tr}}^{\infty} \rangle$ for worms, since the operation would simply reproduce the measured values. Instead, we proceed as follows. For each temperature, we compute the trapping-time statistics for individual worms; we then use the mean value to invert Eq.~\eqref{eq:tautr_fit}, thereby obtaining an effective active force for each worm. The effective active force at fixed temperature is then defined as the average over all worms at that temperature; these values are reported in the inset of Fig.~6E.

Finally, using~\eqref{eq:tautr_fit} with the temperature-averaged effective active force, we obtain the predicted values of $\langle \tau_{\mathrm{tr}}^{\infty} \rangle$ used in Fig.~6F of the main text.

\clearpage
\section{Additional data}
\subsection{Conformational phase space}

We report here additional data regarding the ``conformational'' phase space of confined worms at different values of the temperature.

\begin{figure*}[h]
    \centering
    \includegraphics[width=0.75\linewidth]{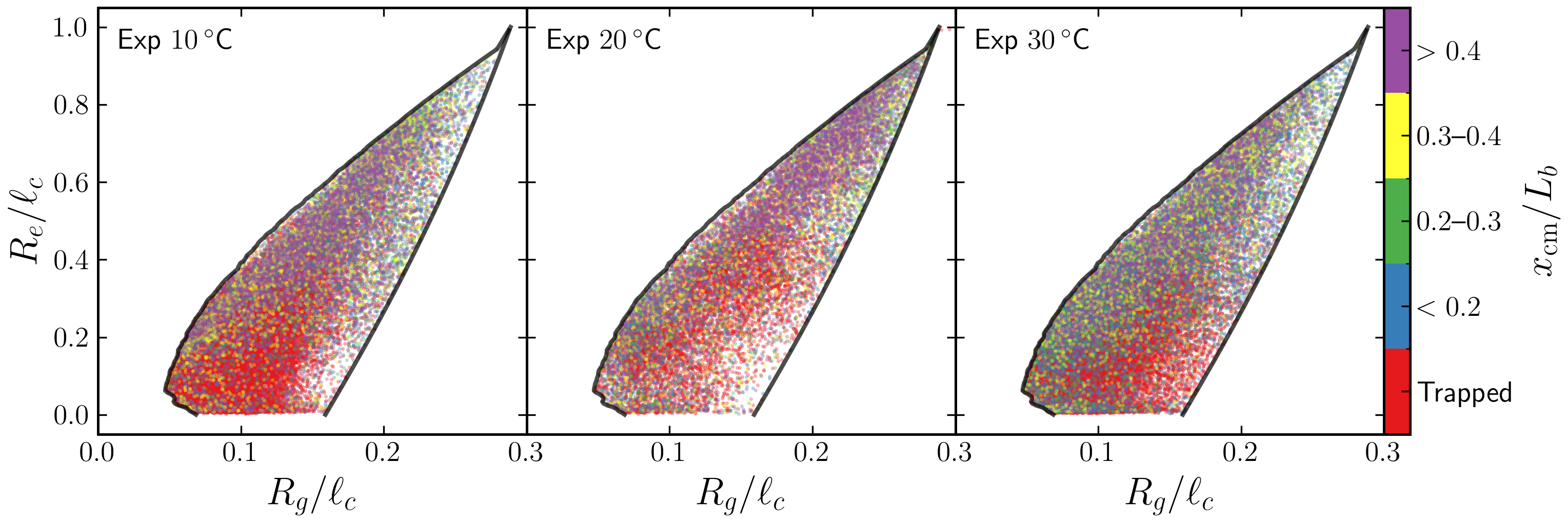}
    \caption{\textbf{Conformational phase space of confined living T. tubifex worms at different values of the temperature.}
    (a) $T=10\,^\circ$C, (b) $T=20\,^\circ$C, and (c) $T=30\,^\circ$C. The normalized end-to-end distance $R_e/\ell_c$ is plotted as a function of the normalized radius of gyration $R_g/\ell_c$, $\ell_c$ being the contour length. Colors indicate the position of the worm’s center of mass along the channel axis, as shown in the color bar; red corresponds to configurations where the center of mass is located inside a cavity.}
    \label{fig:re_rg_vs_T}
\end{figure*}

\begin{figure*}[h]
    \centering
    \includegraphics[width=0.75\linewidth]{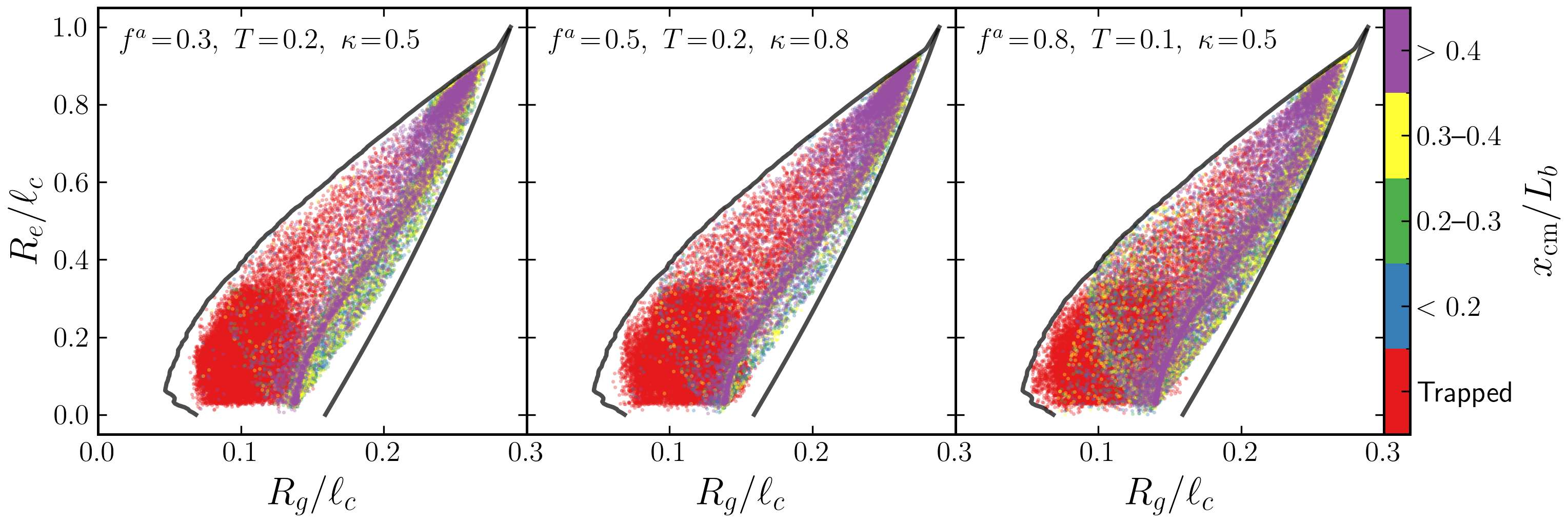}
    \caption{\textbf{Conformational phase space of confined tangentially driven filaments.}
    (a)--(c) Active polymers of length $\ell^*_c = 40\sigma$ with model parameters $(f^a\sigma/u_E, k_BT/u_E, \kappa/u_E) = (0.3, 0.2, 0.5)$, $(0.5, 0.2, 0.8)$, $(0.8, 0.1, 0.5)$, matching the experimental temperatures (a) $10$, (b) $20$, (c) $30\,^\circ$C. Colors indicate the position of the filament’s center of mass along the channel axis, as shown in the color bar; red corresponds to configurations where the center of mass is located inside a cavity.}
    \label{fig:re_rg_vs_T_model}
\end{figure*}

In Fig.~\ref{fig:re_rg_vs_T} we report the normalized end-to-end distance $R_e/\ell_c$ as a function of the radius of gyration $R_g/\ell_c$, with $\ell_c$ the contour length. The different panels refer to data at different value of the temperature. Unlike the plot reported in the main text (Fig.~1,2), here each data point is colored according to the position of the centre of mass of the worm in the connecting channel; red color marks conformations of worms inside one of the cavities. From this plot, one can clearly appreciate that the most compact conformations are attained inside the cavity; conversely, the more extended ones are attained in the channel. This is always true, but it is especially evident at $T=20^{\circ}$C, where it is statistically more probable to observe a worm traversing the channel. At the same time, the fact that the worm is observed more prominently within the chamber at $T=10, 30^{\circ}$C is also evident.

In Fig.~\ref{fig:re_rg_vs_T_model} we report the same representation of the conformational phase space for the active filament model, matching the worms at different values of the temperature, as in the main text. As in the experiments, compact configurations are mostly localized within the cavities, while more extended states are typically found in the connecting channel. Complementing the discussion in the main paper, the over-represented conformations, corresponding to hairpins, are visible also here: this analysis confirms that these conformations are characteristic of the motion through the channel and, specifically, in the centre of the channel, where the filament is fully in the connecting bridge. As mentioned in the main text, these conformations are some of the most difficult to reconstruct.

\subsection{Trapping times}

\begin{figure*}[h]
    \centering
    \includegraphics[width=0.4\textwidth]{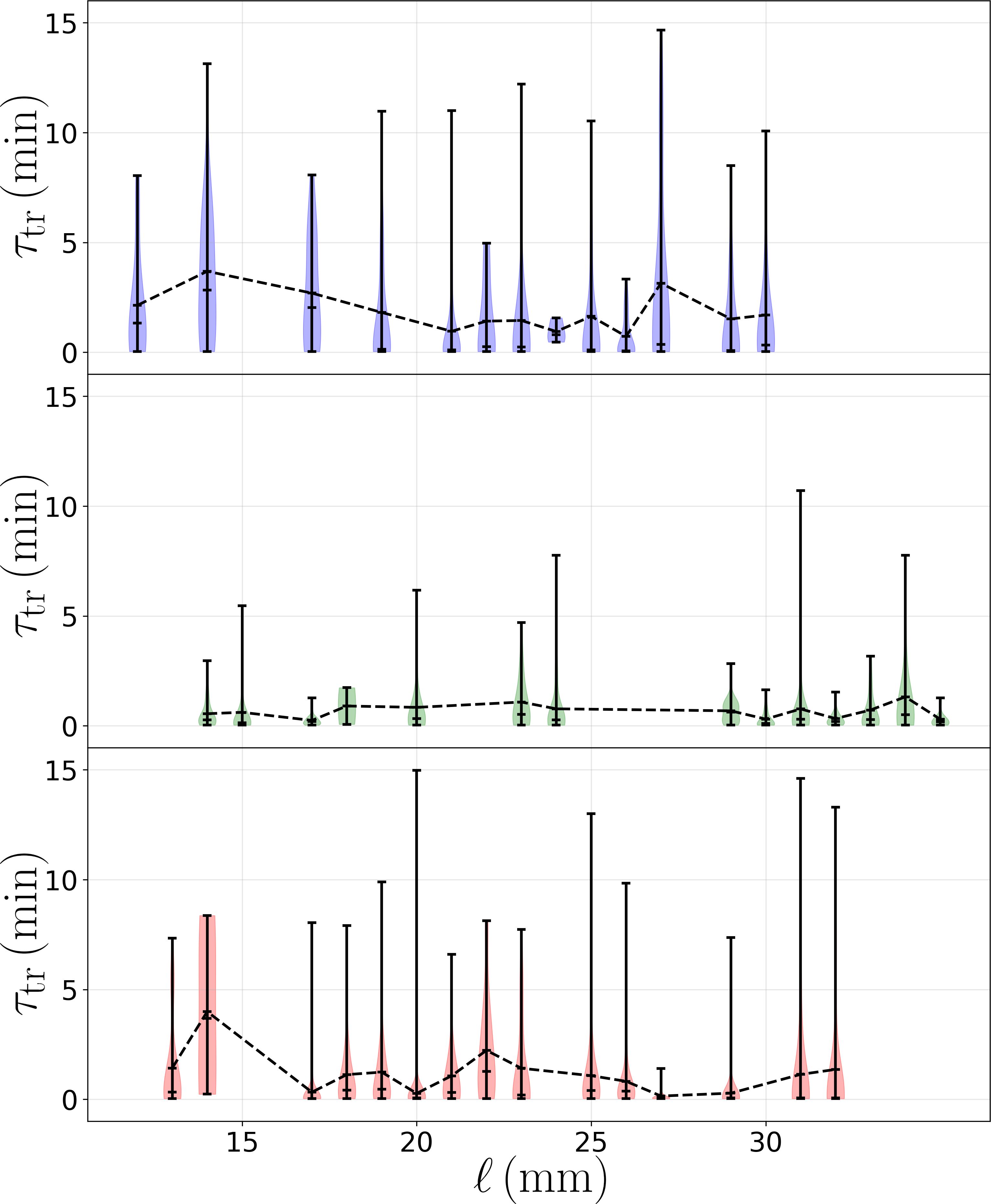}
    \caption{\textbf{Effect of temperature and contour length on trapping time.}
    Trapping times $\tau_{\mathrm{tr}}$ are shown as a function of contour length $\ell_c$ for living filaments at different temperatures: (a) $T=10\,^\circ$C, (b) $T=20\,^\circ$C, and (c) $T=30\,^\circ$C.}
    \label{fig:res_time_suppl}
\end{figure*}

We report here additional data on the trapping times.
We report violin plots of the trapping time as a function of the contour length at different values of the temperature in Fig.~\ref{fig:res_time_suppl}.

On the one hand, we can appreciate that the trapping time at $T=20^{\circ}$C is the lowest of the three. We can also notice that the trapping time remains independent of the contour length of the worms at all values of $T$. \\

We further look at the effect of the bending rigidity on the trapping time. 

\begin{figure*}[h]
     \centering
     \includegraphics[width=0.45\linewidth]{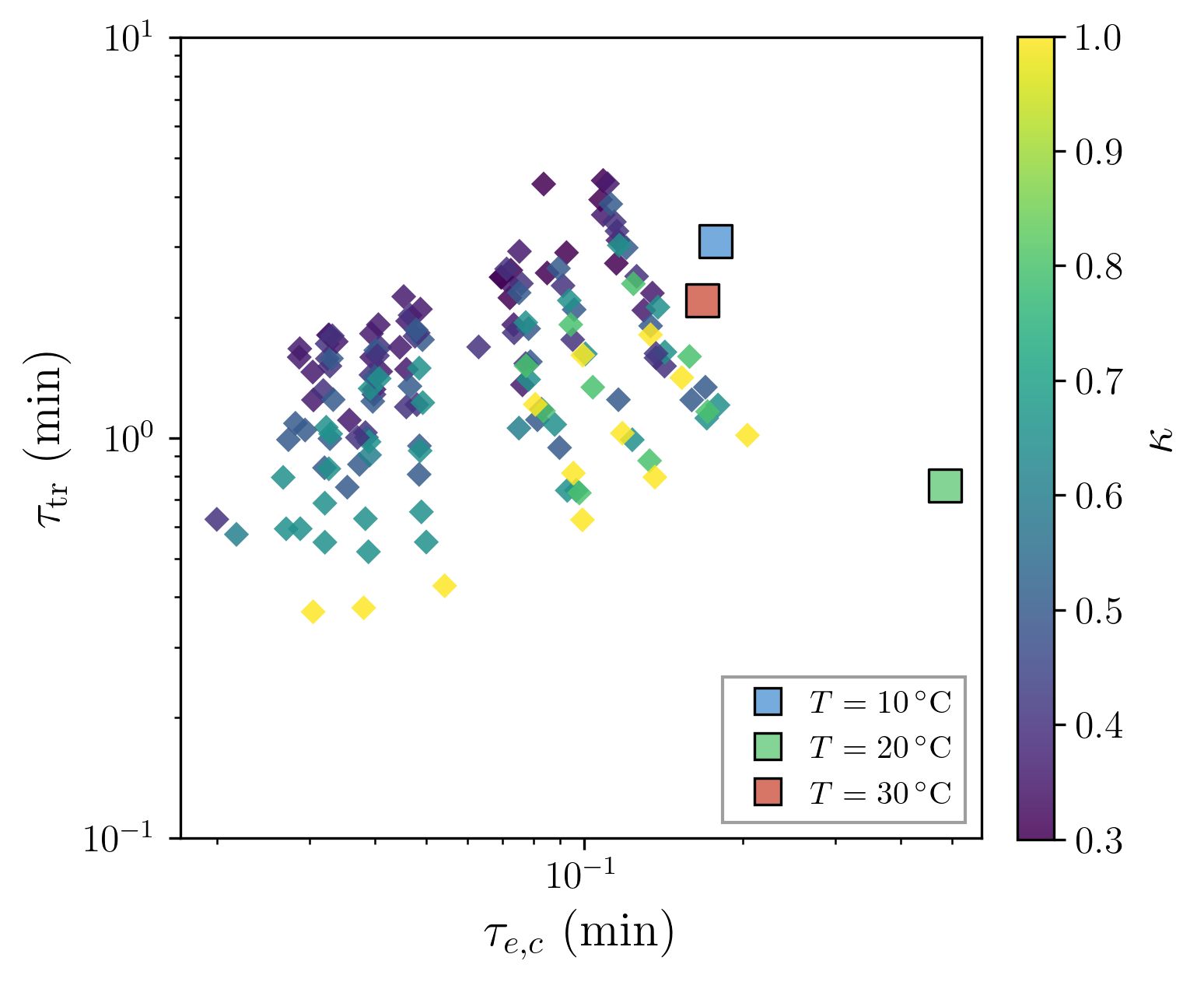}
     \caption{\textbf{Effect of bending stiffness on trapping dynamics.}
Trapping time $\tau_{\mathrm{tr}}$ as a function of the conformational decorrelation time in the cavity, $\tau_{e,c}$, for living worms (squares) and simulated active polymers (diamonds). Colors indicate the polymer bending stiffness $\kappa$. Simulations span $0.3 < \kappa/u_E < 1$, $0.1 < k_B T^*/u_E < 0.3$, and $0.3 < f^a \sigma/u_E < 2$.}
    \label{SupFig:tau_trap_vs_tauec_kappa}
\end{figure*}

In Fig.~\ref{SupFig:tau_trap_vs_tauec_kappa}, we report the trapping time $\tau_{\mathrm{tr}}$ as a function of the conformational decorrelation time in the cavity, $\tau_{e,c}$, for simulated active polymers and living worms; each data point is colored according to its bending rigidity $\kappa$. While Fig.~5A of the main text highlights the role of activity, this representation shows that bending stiffness also contributes to the observed spread in trapping times: stiffer filaments exhibit shorter trapping times. These results further confirm that activity and flexibility jointly control escape dynamics under confinement.

\begin{figure*}[h]
     \centering
     \includegraphics[width=0.45\linewidth]{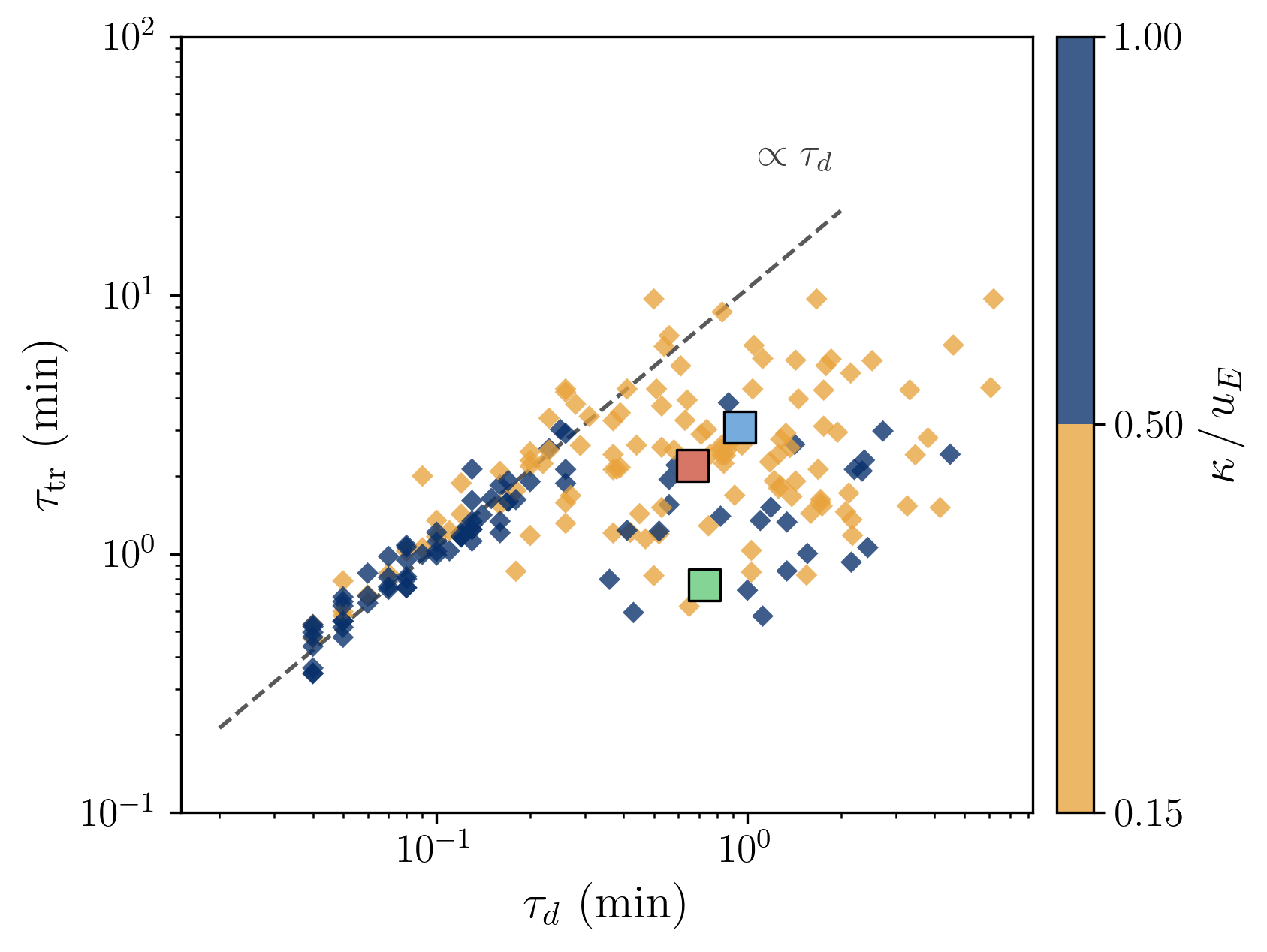}
     \caption{\textbf{Effect of filament stiffness on trapping dynamics.}
     Trapping time $\tau_{\mathrm{tr}}$ as a function of the translational saturation time $\tau_{\mathrm{d}}$ for simulated active polymers. Each point corresponds to a simulation and is colored according to the bending rigidity $\kappa$, separating relatively flexible ($\kappa/u_E < 0.5$, amber) from relatively stiff ($\kappa/u_E \geq 0.5$, navy) filaments. The dashed line indicates the proportional regime $\tau_{\mathrm{tr}} \propto \tau_{\mathrm{d}}$. This regime is predominantly populated by stiffer filaments, consistent with their more persistent, boundary-following dynamics. Colored squares denote experimental measurements for living worms at $T=10$, $20$, and $30\,^{\circ}$C.}
    \label{fig:tau_trap_kappa}
\end{figure*}

In Fig.~\ref{fig:tau_trap_kappa} we report the trapping time $\tau_{\mathrm{tr}}$ as a function of the translational saturation time $\tau_{\mathrm{d}}$ for the simulated active polymers; we color each point according to its bending rigidity $\kappa$. As in the main text, the translational timescale $\tau_{\mathrm{d}}$ is taken as the time at which the head mean square displacement inside the cavity saturates because of the finite cavity size.
We can appreciate that the stiffer filaments ($\kappa/u_E \geq 0.5$) lie predominantly on the proportional line $\tau_{\mathrm{tr}} \propto \tau_{\mathrm{d}}$, while the more flexible ones populate the plateau at large $\tau_{\mathrm{d}}$, supporting the argument in the main text.

\clearpage
\subsection{Conformational entropy}

We report here additional data on the conformational entropy.

\begin{figure*}[h]
     \centering
     \includegraphics[width=0.5\linewidth]{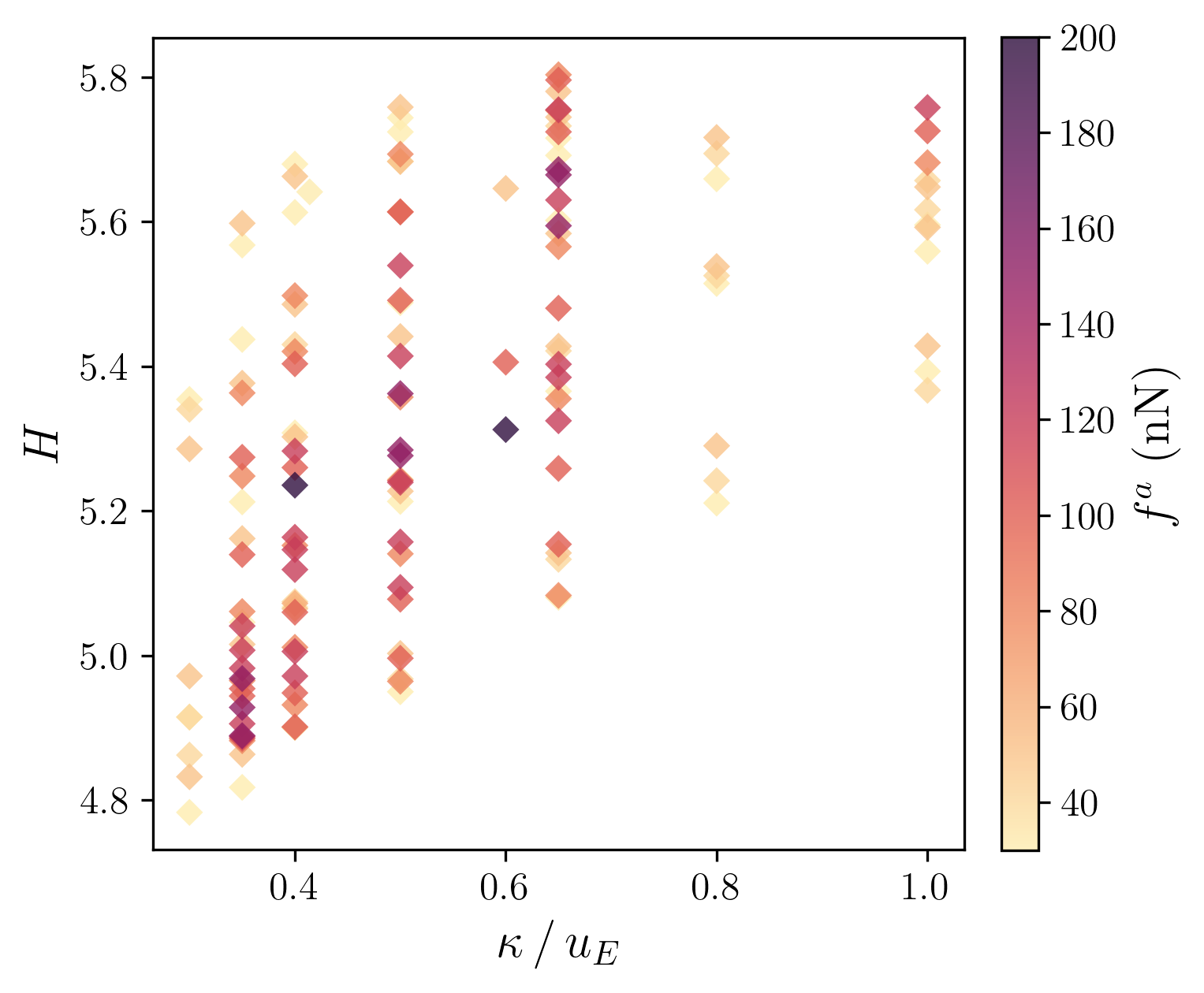}
     \caption{\textbf{Effect of activity and filament stiffness on the conformational entropy.}
     Conformational Shannon entropy $H$ as a function of the bending rigidity $\kappa$ for simulated active polymers. Points are colored according to the active force $f^a$.}
    \label{fig:H_vs_kappa}
\end{figure*}

In Fig.~\ref{fig:H_vs_kappa} we report the conformational Shannon entropy $H$, introduced in the main text, as a function of the bending rigidity $\kappa$ for the simulated active polymers; each point is colored according to the corresponding active force $f^a$. We observe that the entropy depends primarily on filament stiffness and shows only a weak dependence on activity over the range of parameters explored.

\end{document}